\documentclass[aps,twocolumn,prd,superscriptaddress,showpacs,preprintnumbers,amsmath,amssymb]{revtex4}

\usepackage{subfigure}
\usepackage{amsmath,amssymb}
\usepackage{graphicx}
\usepackage{rotating}
\usepackage{bm}
\usepackage{dsfont}
\usepackage{color}
\usepackage{epsfig}

\usepackage{color}

\newcommand{\Tr}{\ensuremath{\operatorname{Tr}}}
\newcommand{\tr}{\ensuremath{\operatorname{tr}}}
\newcommand{\Omegaqq}{\ensuremath{\Omega_{\bar{q}q}}}
\newcommand{\vev}[1]{\ensuremath{\left\langle #1 \right\rangle}}

\def\Eq#1{Eq.~(\ref{#1})}
\def\Fig#1{Fig.~\ref{#1}}

\newcommand{\Phibar}{\ensuremath{\bar{\Phi}}}

\def\dr{{D\!\llap{/}}\,}
\def\Dr{{D\!\llap{/}}\,} 
\def\ipv{\vec{p}\llap{/}}

\def\0#1#2{\frac{#1}{#2}}

\newcommand{\twofigs}{0.4\linewidth}

\newcommand{\pT}{\ensuremath{T_0}}

\graphicspath{
{./}
{../figures/}
}
\begin{document}
\title{The phase structure of the Polyakov--quark-meson model beyond mean field}
\author{Tina Katharina Herbst}
\affiliation{Institut f\"{u}r Physik, Karl-Franzens-Universit\"{a}t,
  A-8010 Graz, Austria}
\author{Jan M. Pawlowski}
\affiliation{Institut f\"ur Theoretische Physik, University of Heidelberg, 
Philosophenweg 16, D-62910 Heidelberg, Germany}
\affiliation{ExtreMe Matter Institute EMMI, GSI Helmholtzzentrum f\"ur 
Schwerionenforschung mbH, Planckstr. 1, D-64291Darmstadt, Germany.}
\author{Bernd-Jochen Schaefer}
\affiliation{Institut f\"{u}r Physik, Karl-Franzens-Universit\"{a}t,
  A-8010 Graz, Austria}



\begin{abstract}
  The Polyakov-extended quark-meson model (PQM) is investigated beyond
  mean-field. This represents an important step towards a fully
  dynamical QCD computation. Both the quantum fluctuations to the
  matter sector and the back-reaction of the matter fluctuations to
  the QCD Yang-Mills sector are included.  Results on the chiral and
  confinement-deconfinement crossover/phase transition lines and the
  location of a possible critical endpoint are presented. Moreover,
  thermodynamic quantities such as the pressure and the quark density
  are discussed.
\end{abstract}

\maketitle

\section{Introduction}
\label{sec:intro}

Driven by the heavy-ion programs at GSI, CERN SPS, RHIC and LHC there
is strong interest in the properties of strongly interacting matter at
extreme temperatures and baryon densities. The access to the phase
diagram of QCD is hampered by the fact that lattice simulations at
finite density suffer from the sign problem \cite{LQCD}. However, in
recent years impressive progress has been made in the understanding of
the phase diagram of QCD within QCD effective models such as the
Polyakov-loop extended Nambu-Jona-Lasinio (PNJL), see
e.g.~\cite{pisarski,Fukushima:2003fw}, and the Polyakov-loop extended
quark-meson (PQM) model, see \cite{Schaefer:2007pw}. In these models
the information about the confining glue sector of the theory is
incorporated in form of a Polyakov-loop potential that is extracted
from pure Yang-Mills lattice simulations. The matter sector of these
models has been studied in detail also beyond the mean-field level by
taking into account the quark-meson quantum fluctuations mostly within
a functional renormalization group (FRG) approach, for recent studies
see \cite{Schaefer:2004en, Braun:2009si}.

However, the most difficult problem within these models is the
question of how to embed the quantum back-reaction of the matter
sector to the gluonic sector. This problem has effectively been
resolved in \cite{Schaefer:2007pw} where the change of $\Lambda_{\rm
  QCD}$, and hence the confinement-deconfinement transition
temperature $T_0$, in the presence of dynamical quarks has been
computed within hard thermal and/or hard dense loop (HTL/HDL)
perturbation theory. This leads to a flavor and chemical potential
dependence of the transition temperature $T_0$, and is a qualitatively
viable procedure: the change of the non-perturbative scale
$\Lambda_{\rm QCD}$ can be very well estimated within perturbation
theory as has been shown and confirmed in many computations at
zero-temperature and finite-temperature QCD. Moreover, this
perturbative estimate in \cite{Schaefer:2007pw} has been confirmed by
first principle QCD computations with the functional renormalization
group at real and imaginary chemical potential in
\cite{Braun:2009gm,BHP}. On the other hand, very recently the
$\mu$-dependence of $T_0$ has been estimated by constraining PNJL
results with those in the statistical model \cite{Fukushima:2010is}.
This links the model parameter $T_0(N_f,\mu)$ under certain weak
assumptions to experimental data about the chemical freeze-out
curve and nicely confirms the theoretical prediction in
\cite{Schaefer:2007pw}.

This shows that even though being QCD effective models, they provide
valuable information about the phase diagram, and in particular help
to exclude certain scenarios. Moreover, the models can be understood
as specific controlled approximations to full dynamical QCD. Most
directly this is realized within the above-mentioned FRG approach put
forward in \cite{Braun:2009gm, BHP, Braun:2008pi, Marhauser:2008fz,
  Kondo:2010ts}.  This link enables us to systematically extend the
PNJL/PQM models towards full dynamical QCD.

In the present work we take an important step towards full dynamical
QCD and present the first computation with fully dynamical matter
sector in the PQM model at finite density. An interesting first FRG
computation at vanishing density has been put forward in
\cite{Skokov:2010wb}. There, however, the back-coupling to the glue
sector has been neglected. Here, we include the back-reaction of the
matter sector as described above. The uncertainty can be estimated by
comparing it to the QCD computation in \cite{Braun:2009gm,BHP}. Within
the present PQM approach beyond mean field we provide results on the
phase boundaries for the confinement-deconfinement and the chiral
transitions as well as on thermodynamic quantities such as the
pressure, quark number density and quark number susceptibility.

The outline of the paper is as follows: in the next section we discuss
the back-reaction of the matter fluctuations to the QCD Yang-Mills
sector in more detail. In Sec.~\ref{sec:pqm}, the Polyakov-loop and
its potential is introduced and coupled to the quark-meson model which
then defines the Polyakov--quark-meson model. At first, the grand
potential of this model is evaluated in mean-field approximation.
Afterwards, the flow equation for the grand potential of this model is
derived in Sec.~\ref{sec:flow} where also the choice of model
parameters is discussed. Sec.~\ref{sec:phase_structure} is devoted to
the phase structure and some thermodynamical applications, in
particular we evaluate the pressure, quark number density and quark
number susceptibility. The phase structure, i.e., the chiral and
confinement-deconfinement phase transition is explored in detail.
Subsequently, the influence of the Polyakov loop on the thermodynamics
is investigated and in Sec.~\ref{sec:conclusion} concluding remarks
are drawn.

\section{The PQM model beyond mean field} 
\label{sec:beyond}

The Polyakov--quark-meson (PQM) model was studied in a mean-field
approximation for the matter sector in \cite{Schaefer:2007pw}. In this
model, the coupling between the pure glue sector, realized by an
effective Polyakov-loop potential $\cal U$, and the quark-meson matter
sector is provided by the fermionic determinant. However, in
\cite{Schaefer:2007pw} an important step beyond the mean-field
approximation was already introduced. More specifically, the pure glue
potential $\cal U$ was adjusted on the basis of phenomenological
arguments: the phase transition temperature $T_0$ in the potential
$\cal U$ relates to the dynamical scale in QCD, $\Lambda_{\rm QCD}$.
If the dynamics of the quarks is switched on, this scale is lowered.
In \cite{Schaefer:2007pw} we have provided a flavor and quark chemical
potential dependent transition temperature $T_0(N_f,\mu)$ which was
estimated from hard thermal and hard dense loop considerations. Here,
we shall elaborate this argument in more detail, also in order to show
that no double counting is involved in the computation and to tighten
the link to first principle QCD. We also stress that the above
phenomenological argument works well at large chemical potential $\mu$
as well as at small chemical potential. For vanishing chemical
potential we also compare our results with a recent dynamical QCD
calculation \cite{Braun:2009gm}. This allows us to qualitatively
adjust the phase transition lines at small chemical potential and
small temperature, and gives us access to information about a possible
critical endpoint as well as the size of a possible quarkyonic matter
region at large chemical potential in the QCD phase diagram as
suggested in \cite{McLerran:2007qj}.

What is then left is to include the quantum and thermal fluctuations
of the matter sector in the presence of the non-trivial Polyakov loop.
This is achieved with functional renormalization group (FRG) methods
and is done in Sec.~\ref{sec:flow}, for reviews see e.g. \cite{FRG,
  Pawlowski:2005xe}. In combination, this allows for a Polyakov-loop
effective model computation which already takes into account the full
dynamics of QCD phenomenologically.

\begin{figure}[ht]
\centerline{\epsfig{file=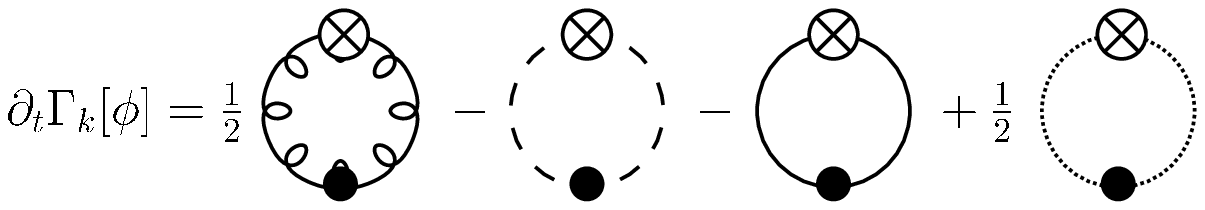,width=8cm}}
\caption{Functional QCD flow for the effective action: The lines denote
  the corresponding full field dependent propagators. Crosses denote
  the cut-off insertion $\partial_t R$. }
\label{fig:funflow}
\end{figure}

We proceed with the explanation of the $N_f$ and $\mu$ dependence of
the transition temperature $T_0$. In \cite{Braun:2009gm} a full
two-flavor QCD computation was put forward for vanishing and
imaginary chemical potential. The related functional RG equation for
the effective action is provided in a simple diagrammatic form in
Fig.~\ref{fig:funflow}.
%
%
The first two loops in Fig.~\ref{fig:funflow} stand for quantum
fluctuations (gluons and ghosts) in the pure glue sector that, e.g.,
generate the Polyakov-loop potential. The third loop represents the
quark fluctuations, and the last term encodes mesonic fluctuations
generated by dynamical hadronization \cite{Gies:2001nw,
  Pawlowski:2005xe, Floerchinger:2009uf}. The crosses denote the
cut-off insertion which restricts the loop momenta to that about the
cut-off scale, and $\phi$ stands for all fields. For more details see
\cite{Braun:2009gm,BHP,Braun:2008pi}. 

We would like to emphasize that the above equation for the effective
action is fully coupled. An important example is the gluonic (and
indirectly the ghost) propagation that is modified in the presence of
dynamical quarks.
%
\begin{figure}[ht]
\centerline{\epsfig{file=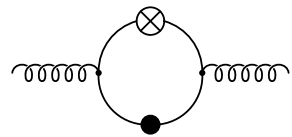,width=4cm}}
\caption{Quark contributions to the flow of the gluon propagator}
\label{fig:quarkloop}
\end{figure}
%
This aspect is visualized in Fig.~\ref{fig:quarkloop} where the quark
contribution to the gluonic propagator flow is shown. In turn, there
are gluonic contributions to quark and meson correlation functions, and
the system is highly non-linear. 

\begin{figure}[ht]
  \centerline{\epsfig{file=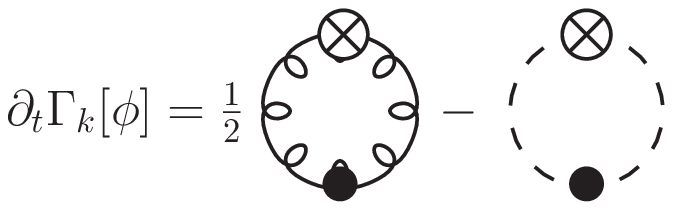,width=5.5cm}}
  \caption{The pure Yang-Mills flow for the effective action.}
  \label{fig:funflowYM}
\end{figure}

Another important aspect is that the pure Yang-Mills (YM)
Polyakov-loop potential can be derived from the pure Yang-Mills flow
as depicted in Fig.~\ref{fig:funflowYM}.
%
%
This simply amounts to dropping the matter loops in
Fig.~\ref{fig:funflow}. The pure Yang-Mills computation has been done
in \cite{Braun:2007bx}, and the related $T_0$ agrees quantitatively
with lattice predictions which are used to adjust the $T_0$ parameter
in the Polyakov-loop potential $\cal U$. These findings also elucidate
that $T_0$ is nothing else but the dynamical YM-scale $\Lambda_{\rm
  YM}$ inherent in the ghost and gluon propagators. This supports the
quantitative accuracy of such an approach.

In turn, the quantum dynamics of the quark-meson model is obtained by
dropping the Yang-Mills diagrams in Fig.~\ref{fig:funflow}, that is
the gluon and ghost loop. This has been studied extensively in the
literature, for recent works see
e.g.~\cite{Schaefer:2004en,Braun:2009si}. In summary, it is this
simple additive structure for the different contributions that allows
to systematically improve quark-meson models towards full QCD within
an FRG approach. Moreover, it also allows us to directly use full QCD
information within the PQM/PNJL models. This reduces the
parameter-dependence of these models, and qualitatively enhances the
predictive power.

An important example for this structure is the Polyakov-loop potential
in these models: the Polyakov-loop potential $\cal U$ in full
dynamical QCD still comes from the Yang-Mills contributions in
Fig.~\ref{fig:funflow}. The related Polyakov-loop potential in QCD can
be computed from the full dynamical QCD computation in
\cite{Braun:2009gm}, and receives a flavor and chemical potential
dependence via Fig.~\ref{fig:quarkloop}. It is clear that the quark
contributions to the gluon propagation (and ghost propagation) shown
in Fig.~\ref{fig:quarkloop} lower the dynamical scale in the gluon and
ghost propagation, and hence $T_0$. In turn, with constant $T_0$
without a flavor and $\mu$-dependence, this contribution is estimated
by the pure Yang-Mills results. We conclude that the hard thermal and
hard dense loop adjustment of $T_0$ is necessary for the full
inclusion of quark effects to the Polyakov-loop potential. In
particular, the above contributions are not included in the fermionic
determinant and thus no double counting is involved.

In summary, the inclusion of the $N_f$ and $\mu$-dependence in $T_0$
as well as the dynamics of the quark-meson sector of the model in the
presence of a non-trivial Polyakov-loop expectation value gives us a
good qualitative control about the full QCD dynamics, and the link is
tightened by comparing $T_0(N_f,\mu)$ with QCD results in
\cite{Braun:2009gm,BHP}. The direct use of the full QCD Polyakov-loop
potential will be discussed elsewhere.

\section{The Polyakov-Quark-Meson Model}
\label{sec:pqm}

The Polyakov-quark-meson model as put forward in
\cite{Schaefer:2007pw} already provides a good approximation of full
QCD at low energies and temperatures and not too high densities. Its
classical Euclidean action with $N_f=2$ light quarks $q=(u,d)$ and
$N_c=3$ color degrees of freedom reads
\begin{eqnarray} \label{eq:pqmmodel}
  S &=& \int d^4 x\Biggl\{\bar{q} \,(\dr + h (\sigma + i \gamma_5
  \vec \tau \vec \pi ))\,q +\frac 1 2 (\partial_\mu \sigma)^2 
  \nonumber \\[1ex]
  && \qquad  + \frac{ 1}{2}
  (\partial_\mu \vec \pi)^2
 +U(\sigma, \vec \pi )  +{\cal U}(\Phi,\bar\Phi)\Biggr\}\ ,   
\end{eqnarray}
where $\dr(\Phi)=\gamma_\mu \partial_\mu-i\,g \gamma_0 A_0(\Phi)$ with
gauge coupling $g$ and $h$ is the Yukawa coupling between mesons and
quarks. The temporal component of the gauge field is linked to the
order parameter of quark confinement (in pure Yang-Mills), the
Polyakov-loop variable $\Phi$,
\begin{equation}
  \label{eq:defphi} 
  \Phi(\vec x) = \frac{1}{N_c} \left\langle \ensuremath{\tr} {\cal P} 
    \exp \left( i\, g\int_0^{\beta} d\tau A_0   
      (\vec x, \tau ) \right)
  \right\rangle\,, 
\end{equation}
where the color trace $\tr$ is in the fundamental representation,
${\cal P}$ denotes path ordering, and $\beta=1/T$, the inverse
temperature. The purely mesonic potential is defined as
\begin{eqnarray} \label{eq:pot}
U(\sigma, \vec \pi ) &=& \frac \lambda 4 (\sigma^2+\vec \pi^2 -v^2)^2
-c\sigma\ \,.
\end{eqnarray}
The isoscalar-scalar $\sigma$ field and the three
isovector-pseudoscalar pion fields $\vec \pi$ together form a chiral
vector field $\vec \phi$. Without the explicit symmetry breaking term
$c$ in the mesonic potential the Lagrangian is invariant under global
chiral $SU(2)_L\times SU(2)_R$ rotations.  

The parameters of the model defined in \Eq{eq:pqmmodel} are fixed at
the low energy physics at vanishing temperature.

\subsection{Polyakov-loop potential}
\label{sec:polypots}

In recent years, several Polyakov-loop potentials have been
proposed~\cite{Roessner:2006xn, Fukushima:2008wg}, for comparison see
\cite{Schaefer:2009ui}. Their functional form is motivated by the
underlying QCD symmetries in the pure gauge limit and they all
reproduce a first-order transition at $T_c\sim 270$ MeV for $N_c=3$
colors in this limit.

In this work we use a polynomial ansatz for the Polyakov-loop
potential
\begin{multline}
\label{eq:upoly}
\frac{\mathcal{U}_{\text{poly}}}{T^{4}}= -\frac{b_2(T)}{2}\Phi\Phibar 
- \frac{b_3}{6}\left(\Phi^{3}+\Phibar^{3}\right)+ \frac{b_4}{4} \left(\Phi\Phibar\right)^{2}\ ,
\end{multline}
with the temperature-dependent coefficient
\begin{equation}
\label{eq:upolypara}
  b_2(T) =  a_0 + a_1 \left(\frac{T_0}{T}\right) + a_2
  \left(\frac{T_0}{T}\right)^2 + a_3 \left(\frac{T_0}{T}\right)^3. 
\end{equation}

The parameters of Eqs.~(\ref{eq:upoly}) and (\ref{eq:upolypara}) are
fitted to the pure gauge lattice data with
\begin{equation}
a_0 = 6.75\ ,\  a_1= -1.95\ , \ a_2 = 2.625\ ,\ a_3 = -7.44\ 
\end{equation}
and
\begin{equation}
 b_3=0.75\ ,\quad b_4 = 7.5 \ .
\end{equation}
The above potential reproduces well the equation of state and the
Polyakov-loop expectation value, in particular around the transition
\cite{Schaefer:2009ui}. The parameter $T_0 = 270$ MeV corresponds to
the transition temperature in the pure YM theory.

As motivated in the previous section, it is left to fix $T_0$ in the
presence of dynamical quarks. Since it is directly linked to the
dynamical mass-scale $\Lambda_{\rm QCD}$ this parameter necessarily
has a flavor and chemical potential dependence in full dynamical QCD
and $T_0\to T_0(N_f,\mu)$.

In the present work we use perturbative relations for fixing the
relative scales \cite{Schaefer:2007pw,Braun:2009ns}. The results
compare well to the full $N_f=2$ QCD computation in the chiral limit
in \cite{Braun:2009gm,BHP}. The latter thus allows for an error
estimate of the present procedure. The one-loop $\beta$-function of
QCD with massless quarks is given by
\begin{eqnarray}
\label{eq:beta}
\beta(\alpha)=-b \alpha^2\,,
\end{eqnarray}
with the coefficient
\begin{eqnarray}\label{eq:coeffs}
  b(N_f)&=& \0{1}{6 \pi} (11 N_c-2 N_f)\,. 
\label{eq:coeffs2}\end{eqnarray}
Here, we have assumed a RG scheme that minimizes (part of) the
higher-order effects. At leading order the corresponding gauge coupling
is given by
\begin{eqnarray}
  \label{eq:coupling} 
  \alpha(p)=\0{\alpha_0}{1+\alpha_0 b(N_f) \ln
    (p/\Lambda)}+O(\alpha_0^2) \,,
\end{eqnarray}
with $\alpha_0=\alpha(\Lambda)$ at some UV-scale $\Lambda$. The scale
$\Lambda_{\rm QCD}=\Lambda \exp(-1/(\alpha_0 b))$ corresponds to the
Landau pole of \Eq{eq:coupling}.

The temperature dependence of the coupling is also governed by
\Eq{eq:coupling} with the identification $p\sim T$. This yields the
relation \cite{Schaefer:2007pw}
\begin{eqnarray}\label{eq:relation} 
  T_0(N_f)=\hat T e^{ -1/(\alpha_0 b(N_f))} \,,
\end{eqnarray}
where $\hat T$ and $\alpha_0$ are free parameters. \Eq{eq:relation}
allows us to determine the $N_f$-dependence of the critical
temperature $T_0(N_f)$. Analogously to \cite{Schaefer:2007pw} we
choose $\hat T$ to be the $\tau$-scale, $\hat T=T_\tau=1.77$ GeV. This
constitutes a reasonable UV scale for the mean-field model. Then the
pure Yang-Mills input, $T_0(N_f=0)=270$ MeV, leads to
$\alpha_0=0.304$. In the present work we shall stick to these
values. In addition to the arguments given in \cite{Schaefer:2007pw},
the ratio $\pT/T_{\chi}$ in the chiral limit compares well with that
computed in the full two-flavor QCD calculation in
\cite{Braun:2009gm}. Table~\ref{tab:critt} summarizes the
$N_f$-dependent critical temperature $T_0$ in the Polyakov-loop
potential for massless flavors:
\begin{table}[h!]
  \begin{tabular}{c||@{\hspace{2mm}}c@{\hspace{2mm}}|@{\hspace{2mm}}
      c@{\hspace{2mm}}|@{\hspace{2mm}}c@{\hspace{2mm}}|@{\hspace{2mm}}
      c@{\hspace{2mm}}|@{\hspace{2mm}}c@{\hspace{2mm}}}
    $N_f$ & $0$ & $1$ & $2$ & $2+1$ & $3$  \\
    \hline
    \hline
    $T_0$ [MeV] & 270 & 240 & 208 & 187 & 178 \\
  \end{tabular}
  \caption{\label{tab:critt} Critical Polyakov-loop temperature $T_0$ for
    $N_f$ massless flavors.}
\end{table}

Massive flavors lead to suppression factors of the order
$T_0^2/(T_0^2+m^2)$ in the $\beta$-function. For $2+1$ flavors and a
current strange quark mass $m_s\approx 150$ MeV we obtain
$T_0(2+1)=187$ MeV. We estimate the systematic error for $T_0(N_f)$
being of the order ${}^{+15}_{-20}$ MeV related to the scale matching
of the present PQM computation with the QCD computation in the chiral
limit in \cite{Braun:2009gm}. Note, however, that the link to QCD
qualitatively improves the error estimate in comparison to the
estimate done in \cite{Schaefer:2007pw}.

As argued in the last section, in addition to the flavor dependence
of $\pT$ we introduce a chemical potential dependence via a
$\mu$-dependent running coupling $b$, which should push the
confinement-deconfinement transition temperature down close to the
chiral transition line. This can be achieved by defining
\begin{equation}
  \label{eq:t0mu}
  \pT(N_f, \mu) = T_{\tau} e^{-1/(\alpha_0 b(N_f, \mu))}
\end{equation}
with
\begin{equation}
 b(N_f, \mu) = b(N_f) - b_{\mu}\frac{\mu^2}{(\hat{\gamma}\ T_{\tau})^2}\ .
\end{equation}
The factor $\hat{\gamma}$ is a parameter governing the curvature of
$\pT(\mu)$ and $b_{\mu}\simeq \frac{16}{\pi}N_f$ as in
\cite{Schaefer:2007pw}. As for the $N_f$-dependence the
$\mu$-dependence in \Eq{eq:t0mu} compares well to that found in QCD
\cite{Braun:2009gm, BHP}. Based on the results there we estimate the
systematic error with $0.7 \lesssim\hat \gamma \lesssim 1$, and we
shall investigate the $\hat\gamma$-dependence of our results in
Sec.~\ref{sec:phase_structure}.

\subsection{Grand Potential in Mean-Field Approximation}
\label{sec:meanfield}

All thermodynamic properties of the PQM model follow from the grand
potential. It is a function of the temperature and one quark chemical
potential since we consider the $SU(2)_f$-symmetric case in this work
and set $\mu \equiv \mu_u=\mu_d$.

In the mean-field approximation certain quantum and thermal
fluctuations in the path integral representation of the grand
potential are neglected. The mesonic quantum fields are replaced by
their corresponding classical expectation values and only the
integration over the quark loop is performed which is modified by
constant gluon background fields in the PQM model \cite{Marko:2010cd}.
The final potential in mean-field approximation reads
\begin{equation}\label{eq:OmegaMF}
  \Omega^{\ }_{\rm MF} = \Omegaqq({\sigma},\Phi,\bar\Phi) +
  U({\sigma}, 0 )  +{\cal U}(\Phi,\bar\Phi) 
\end{equation}
and consists of the quark contribution including the Polyakov-loop
variables
\begin{eqnarray}
\Omegaqq & =& -2N_fT\int\dfrac{d^3p}{(2\pi)^3} \left \lbrace \ln
  \left[1+3(\Phi+\Phibar e^{-(E_p-\mu)/T}) \right.\right.\nonumber\\
&& \left. \times e^{-(E_p-\mu)/T}+e^{-3(E_p-\mu)/T} \right]\nonumber \\
&& +\ln \left[1+3(\Phibar+\Phi e^{-(E_p+\mu)/T})e^{-(E_p+\mu)/T}\right.\nonumber \\
&& +\left.\left.e^{-3(E_p+\mu)/T}\right] \right\rbrace\ ,
\end{eqnarray}
with the quark/antiquark single-quasiparticle energies
$E_p = \sqrt{\vec{p}^2+m^2_q}$ and the constituent quark mass
$m_q=h \sigma$. The purely mesonic potential $U$ is given by
\Eq{eq:pot} and the effective Polyakov-loop potential $\cal U$, e.g.,
by \Eq{eq:upoly}. Details of the potential derivation can be found in
\cite{Schaefer:2007pw}. The quark contribution involves a divergent
vacuum term which can be regularized. As shown in \cite{Nakano:2009ps,
  Skokov:2010sf} this term is important and modifies the underlying
thermodynamics. Since this term upgrades the standard mean-field
approximation it is neglected here whereas it is included in
the full RG approach.

The solution of the corresponding equations of motion are obtained by
minimizing the thermodynamic potential with respect to the three mean
fields $\sigma$, $\Phi$ and $\bar \Phi$, i.e.,
\begin{equation}
\label{eq:eom}
\left.\frac{ \partial \Omega_{\rm MF}}{\partial \sigma} =\frac{ \partial
  \Omega_{\rm MF}}{\partial \Phi}=\frac{ \partial \Omega_{\rm MF}}{\partial \bar
  \Phi}\right|_{\sigma=\vev{\sigma},\ \Phi = \vev{\Phi},\ \Phibar=\vev{\Phibar}} = 
0 \,.
\end{equation}
The solutions to \Eq{eq:eom} provide the chiral $\vev \sigma$ and
Polyakov-loop expectation values $\vev \Phi$ and $\vev {\bar \Phi}$ as
functions of the temperature and quark chemical potential.

\section{Flow equation}
\label{sec:flow}

The non-perturbative FRG method has a wide range of applicability. In
the context of equilibrium statistical physics it represents a very
efficient way to describe critical phenomena and in particular phase
transitions. One particular formulation of RG flows is based on the
concept of the effective average action $\Gamma_k$ where $k$ denotes a
RG momentum scale, for reviews see e.g.~\cite{FRG, Pawlowski:2005xe}.
For a system with bosonic ($\varphi$) and fermionic fields ($\psi$),
the variation of $\Gamma_k$ with the RG scale ($t = \ln k$) is
governed by the flow equation
\begin{eqnarray} \label{eq:flowQM}
\partial_t \Gamma_k [\varphi,\psi] & = &\frac{ 1}{2} \Tr \left\lbrace
  \partial_t R_{k, B} \left( \Gamma^{(2,0)}_k[\varphi, \psi]+R_{k,
      B}\right)^{-1}\right\rbrace \nonumber\\ 
&& -  \Tr \left\lbrace
  \partial_t R_{k, F} \left( \Gamma^{(0,2)}_k[\varphi, \psi]+R_{k,
      F}\right)^{-1}\right\rbrace\nonumber \\ 
\end{eqnarray}
where $\Gamma_k^{(i,j)}$ denotes the $i$th ($j$th) derivative of
$\Gamma_k$ with respect to the $\varphi$ ($\psi$) fields. The trace
involves a $d$-dimensional momentum integration and a summation over
all inner spaces (flavor, color and/or Dirac). The regulator functions
$R_{k, i}$ and their derivatives implement Wilson's idea of
integrating successively over narrow momentum shells and ensure that
the flow equation is both infrared and ultraviolet finite. We employ
the optimized regulator functions \cite{Litim:2001up} which depend
only on the spatial components of the momentum and for bosonic fields
is given by
\begin{equation}
 R_{k, B} = \left( k^2 - \vec{p}^2\right) \Theta\left(1-\frac{\vec{p}^2}{k^2}\right) 
\end{equation}
whereas the fermionic regulator for the PQM model reads 
\begin{equation}
  R_{k, F} = i\ipv \left(\sqrt{\frac{k^2}{\vec{p}^2}}-1\right)\Theta
  \left(1-\frac{\vec{p}^2}{k^2}\right)\ .
\end{equation}
Further details and the finite temperature and density generalization
of the flow equation can be found, e.g., in Refs.~\cite{Liao:1995gt,
  Schaefer:1999em, Schaefer:2004en, Braun:2003ii, Litim:2001ky,
  Litim:2006ag}.

In the PQM model with the classical action, \Eq{eq:pqmmodel}, the flow
\Eq{eq:flowQM} for the quark-meson sector also depends on the Polyakov
loop via the Dirac term. Here we will take into account the full
dependence of the quark and meson fluctuations on a general constant
$\Phi,\bar\Phi$ background. In summary, this corresponds to the
truncation
\begin{eqnarray} \label{eq:initialpqm} \Gamma_k &=& \int d^4x \left\{
    \bar{\psi} \left(\Dr +\mu \gamma_0+ ih (\sigma + i
      \gamma_5 \vec \tau \vec \pi )\right)\psi \right.\\[1ex]
  && \left.+\frac 1 2 (\partial_\mu \sigma)^2+ \frac{ 1}{2}
    (\partial_\mu \vec \pi)^2 +{\Omega}_k[\sigma,\vec
    \pi,\Phi,\bar\Phi]\right\}\ , 
\nonumber 
\end{eqnarray} 
where ${\Omega}_k[\sigma,\vec \pi,\Phi,\bar\Phi]$ is the full
effective or grand potential in a general background $\Phi$ and
$\bar\Phi$. It is the quantum analogue of the mean-field potential
$\Omega_{\rm MF}$, given in \Eq{eq:OmegaMF} and, in the present
truncation, carries the full $k$-dependence of the effective action
$\Gamma_k$. The approximation \Eq{eq:initialpqm} has also been used
for the matter sector in \cite{Braun:2009gm} and in the first PQM
study beyond mean field at vanishing density \cite{Skokov:2010wb}.
Finally, we are interested in $\Omega_{k=0}$ evaluated at the
solutions of the quantum equations of motion. This corresponds to the
thermodynamic potential where from all temperature, chemical potential
and field derivatives follow. The spatial gauge field is set to zero
while we keep the temporal component as a constant mean-field
background. With this truncation the flow equation for the effective
potential in leading order derivative expansion reads
\begin{widetext}
  \begin{eqnarray}
    \label{eq:flow}
    \partial_t \Omega_k[\sigma,\vec
    \pi,\Phi,\bar\Phi] & = &
    \frac{k^5}{12\pi^2}\left[
      \frac{3}{E_{\pi}}\text{coth}\left(\frac{E_{\pi}}{2T}
      \right)+
      \frac{1}{E_{\sigma}}\text{coth}\left(\frac{E_{\sigma}}{2T}
      \right)
      -\frac{2\nu_q}{E_q}\left\{
        1-N_{q}(T,\mu;\Phi,\bar{\Phi})-N_{\bar q}(T,\mu;\Phi,\bar{\Phi}) \right\} \right] 
  \end{eqnarray}
  with the Polyakov-loop enhanced quark/antiquark occupation
  numbers
  \begin{eqnarray}
    N_{q}(T,\mu;\Phi,\bar{\Phi}) \!&\! = \!&\! \dfrac{1+2\bar{\Phi} e^{(E_q-\mu)/T}+\Phi
      e^{2(E_q-\mu)/T}}{1+3\bar{\Phi}
      e^{(E_q-\mu)/T}+3\Phi e^{2(E_q-\mu)/T}+e^{3(E_q-\mu)/T}}
    \quad  \text{and} \quad 
    N_{\bar q}(T,\mu;\Phi,\bar{\Phi})  \equiv  N_{q}(T,-\mu;\bar{\Phi},\Phi)\
    ,\   
  \end{eqnarray}
\end{widetext}
see also the QCD study \cite{Braun:2009gm} at vanishing and imaginary
chemical potential, and the PQM study \cite{Skokov:2010wb} at
vanishing chemical potential. In the present work we explore the full
phase diagram at real chemical potential. The number of internal quark
degrees of freedom is denoted by $\nu_q = 2N_c N_f= 12$. This equation
describes the flow of the full quark and mesonic subsystem modified by
a constant background field. The grand potential also depends on the
Polyakov-loop variables and the expectation value of the square of the
chiral 4-component field $\phi^2 = \sigma^2 + \vec \pi^2$ which
coincides with $\vev \sigma^2$ since $\vev {\vec \pi}^2=0$.

\begin{figure*}[t]
  \subfigure{\includegraphics[width=\twofigs]{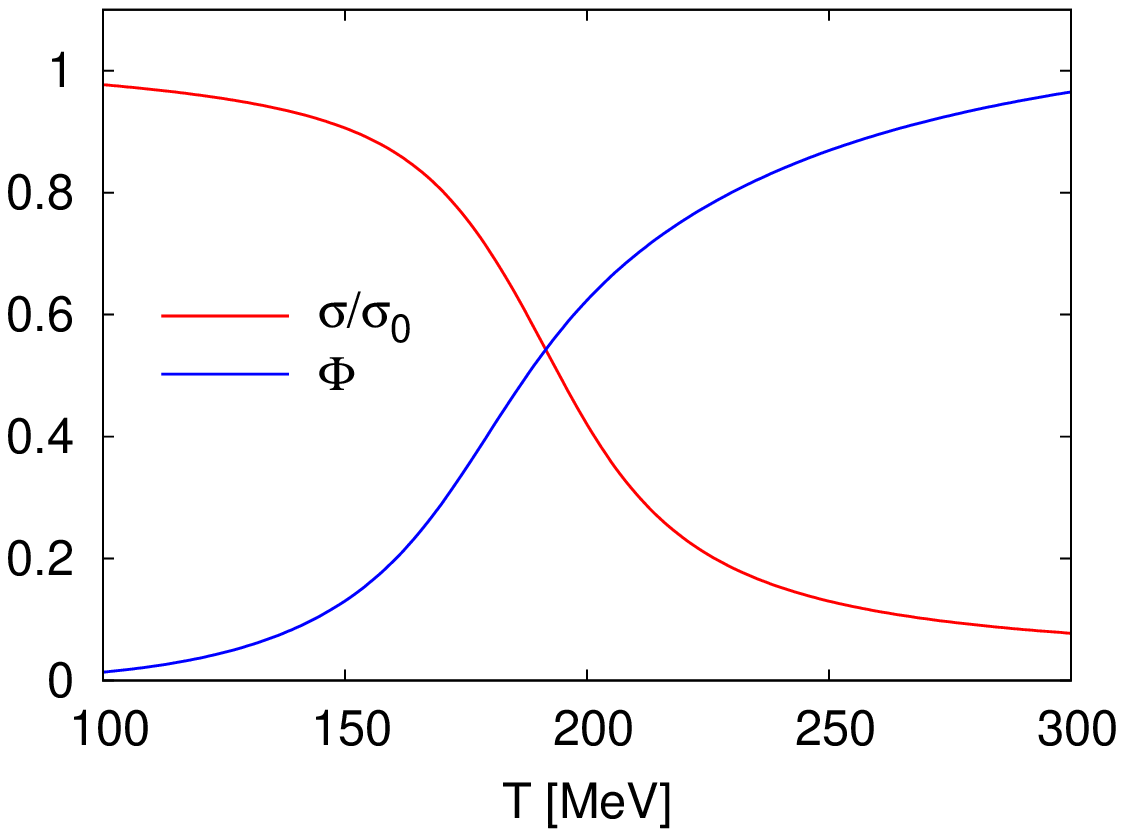}\label{fig:208_orderparams}}
  \subfigure{\includegraphics[width=\twofigs]{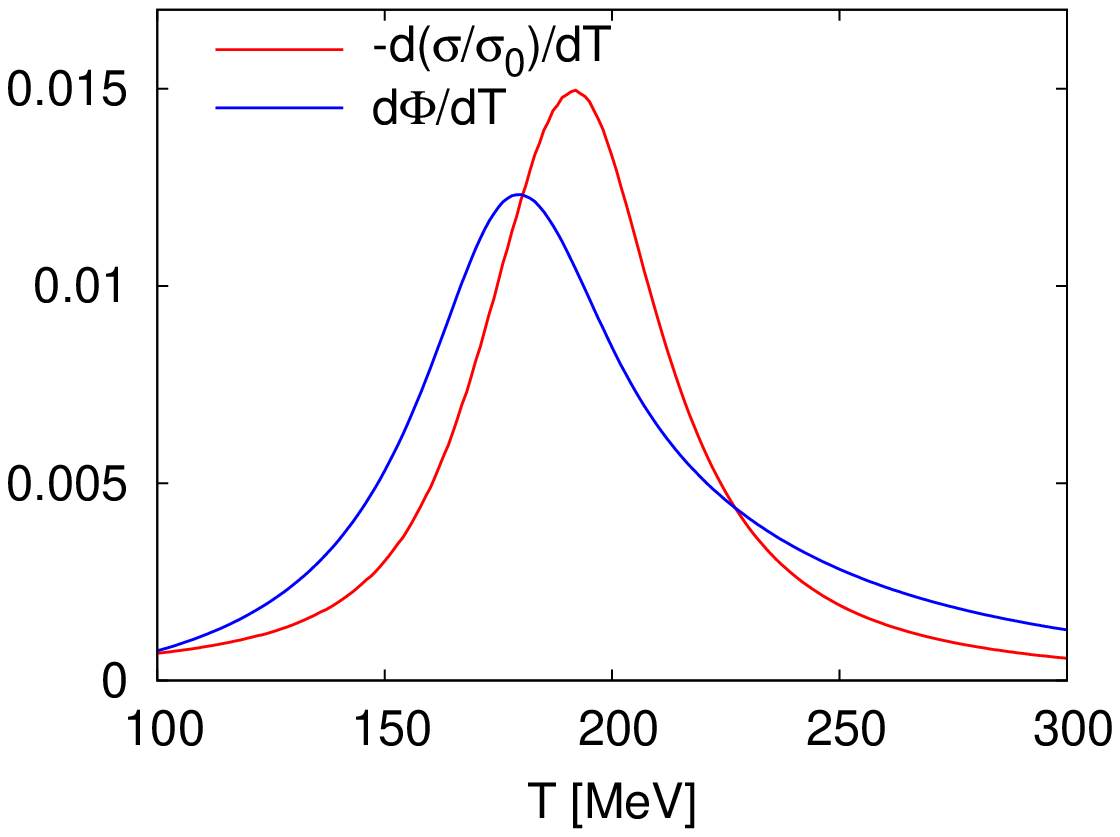}\label{fig:208_Tderivs}}
  \caption{\label{fig:vev208} The normalized chiral condensate
    $\sigma/\sigma_0$ and the Polyakov-loop variable $\Phi$ (left
    panel) and the corresponding temperature derivatives (right panel)
    as a function of the temperature for vanishing chemical potential.
    The Polyakov-loop parameter $T_0=208$ MeV encodes the
    back-reaction to the glue sector of the quark dynamics in the
    presence of two flavors.}
\end{figure*}

The quasi-particle energies for $i=q,\sigma,\pi$ are given by
$E_i= \sqrt{k^2+m_{i}^2}$ with the corresponding squared constituent
quark and meson masses respectively
\begin{equation}
  m_q^2  =  h^2\phi^2\ ,\quad
  m_{\sigma}^2  =  2\Omega'_k+4\phi^2\Omega''_k\ ,\quad
  m_{\pi}^2  =  2\Omega'_k\ .
\end{equation}
Primes denote the $\phi^2$-derivative of the grand potential, i.e.,
$\Omega'_k := \partial \Omega_k /\partial \phi^2$.

In the limit of vanishing background fields, i.e., when
$\Phi,\bar{\Phi}\rightarrow1$, the extended occupation numbers
simplify to the usual Fermi-Dirac distribution functions for quarks
and antiquarks
\begin{eqnarray}
  N_{q}(T,\mu;1,1) & = &
  \dfrac{1}{1+\exp((E_q-\mu)/T)}\ ,\\ 
  N_{\bar q}(T,\mu;1,1) & = &
  \dfrac{1}{1+\exp((E_q+\mu)/T)}\ ,
\end{eqnarray}
and the flow of the quark-meson model is recovered
\cite{QM_jungnickel, Schaefer:2004en}.

The flow equation (\ref{eq:flow}) constitutes a set of coupled, highly
non-linear partial differential equations that cannot be solved with
analytical methods since the right hand side of (\ref{eq:flow})
depends on derivatives of the unknown potential $\Omega_k$. For the
sake of full quantitative precision and in order to facilitate the
access to the potential first-order phase transition region in the QCD
phase diagram we do not resort to Taylor expansions but rather compute
the full numerical solution of the effective potential
$\Omega_k[\sigma,\vec \pi,\Phi,\bar\Phi]$. This extends the analyses
of the full effective potential in the quark-meson model, see
\cite{Schaefer:2004en}, to the present case. The solution of
\Eq{eq:flow} yields the thermodynamic potential
$\Omega_{\rm eff}[\sigma,\vec
\pi,\Phi,\bar\Phi]\equiv\Omega_{k=0}[\sigma,\vec \pi,\Phi,\bar\Phi]$
for the Polyakov-extended quark-meson model in the infrared. The
expectation values of the fields are determined by the quantum
equations of motion,
\begin{equation}
\label{eq:qeom}
\left.\frac{ \partial \Omega_{\rm eff}}{\partial \sigma} =\frac{ \partial
    \Omega_{\rm eff}}{\partial \Phi}=\frac{ \partial \Omega_{\rm eff}}{\partial \bar
    \Phi}\right|_{\sigma=\vev{\sigma},\ \Phi = \vev{\Phi},\ \Phibar=\vev{\Phibar}} = 
0 \,,
\end{equation}
in analogy to the mean-field analysis in \Eq{eq:eom}. 

A full study including algorithmic differentiation (AD) techniques
\cite{Wagner:2009pm} will be presented elsewhere \cite{HPSW}.  In the
present work we shall simply evaluate the effective potential
$\Omega_k$ on the solutions $\Phi(\sigma),\bar\Phi(\sigma)$ of the
mean field EoMs. We shall argue that this is already a quantitatively
reliable approximation to the full solution: note that the present
truncation introduces a cut-off, and hence implicitly a
momentum-dependent fermion propagator. For large temperatures $T$ in
the deconfined regime the fermion propagator is cut-off by the
Matsubara mass $\pi T$. At small temperatures the theory is chirally
broken and the fermion propagator exhibits a mass of the order of the
chiral scale $\sim 360$ MeV. Moreover, at sufficiently large densities
the fermionic fluctuations are already well-described by the one loop
determinant. In summary, the fluctuation dependence of the Polyakov
loop can be treated as a perturbation in the whole phase diagram.
Accordingly, the back-reaction of the Polyakov loop beyond mean field
to the matter fluctuations is suppressed for all $T$ and $\mu$.
Indeed, in comparison to the full solution the minimum for $\sigma$
varies within $\pm 3$ MeV whereas $\Phi,\bar\Phi$ are naturally more
sensitive to this approximation and vary within $\pm 20$ MeV. Note,
that the above structure provides further non-trivial reliability for
the Polyakov-loop extended models. Seen as an expansion towards full
QCD they partially allow for perturbative arguments about the
mean-field analysis. The full computation and a more detailed analysis
of this structure will be provided in \cite{HPSW}. The above structure
also emphasizes the crucial input: the back-reaction of the matter
sector to the Polyakov loop via $T_0(N_f,\mu)$.

It remains to determine the initial effective action
$\Gamma_{\Lambda}$, or, more precisely, the effective potential
$\Omega_\Lambda$ at the initial scale $\Lambda=950$ MeV in the UV.
First of all, consistency with the flow \Eq{eq:flow} leads to a
(relevant) term originated in the fermionic loop,
$\Omega^\infty_\Lambda[\sigma,\vec \pi,\Phi,\bar\Phi]$, see
\Eq{eq:uv_omega}. This term is relevant for the correct thermodynamics
and also includes fermionic vacuum fluctuations. The field-dependent
part of $\Omega_\Lambda$ consists of a sum of the quark-meson
potential \Eq{eq:pot} and the external glue input, the Yang-Mills
Polyakov-loop potential $\cal U$. This yields
\begin{equation}\label{eq:OmegaL}
  \Omega_\Lambda[\sigma,\vec
  \pi,\Phi,\bar\Phi] = U(\sigma,\vec \pi)+{\cal U}(\Phi,\bar\Phi)
  +\Omega^\infty_\Lambda[\sigma,\vec \pi,\Phi,\bar\Phi]\,.
\end{equation}

\begin{figure*}
  \subfigure{\includegraphics[width=\twofigs]{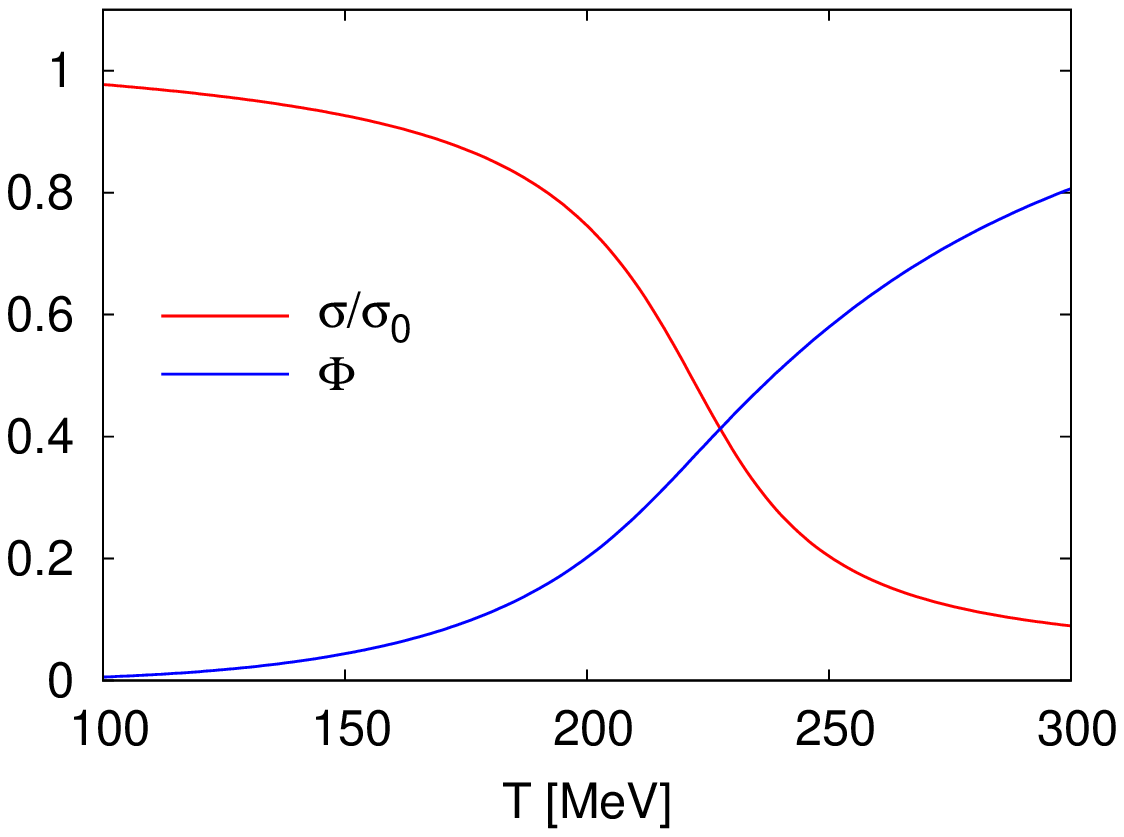}\label{fig:270_orderparams}} 
  \subfigure{\includegraphics[width=\twofigs]{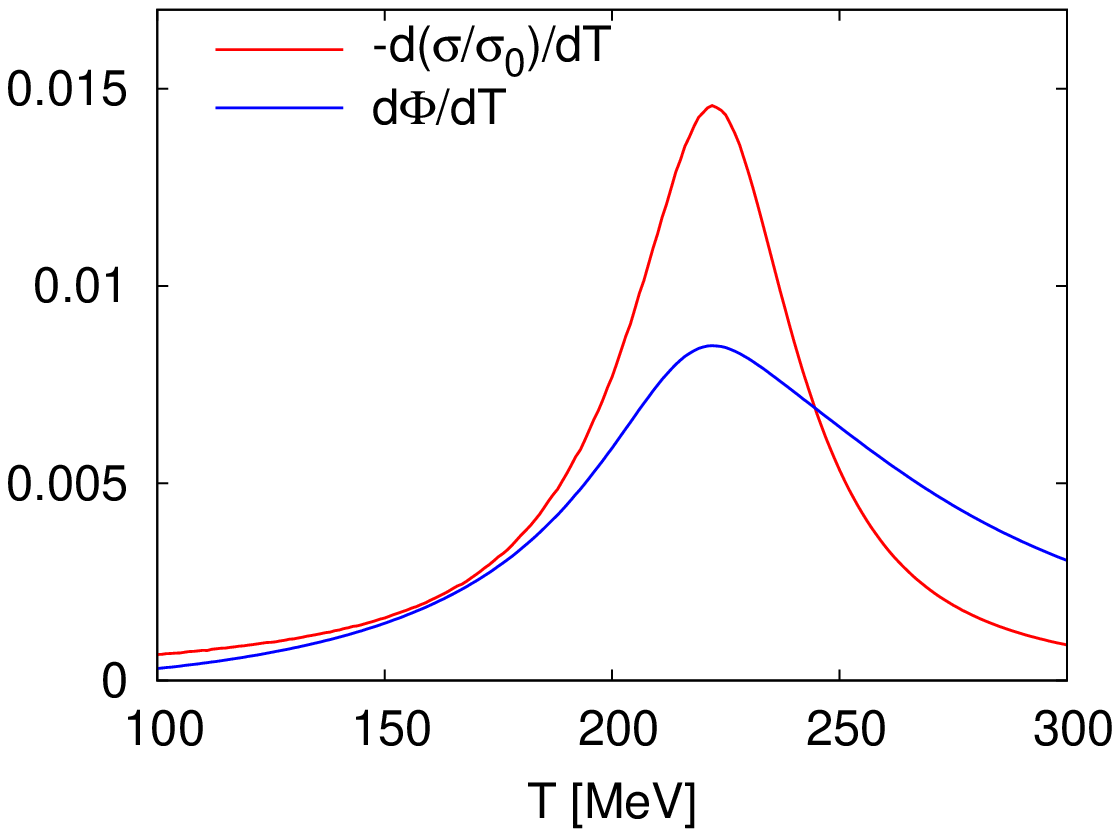}\label{fig:270_Tderivs}}
  \caption{\label{fig:vev270} The normalized chiral condensate
    $\sigma/\sigma_0$ and the Polyakov-loop variable $\Phi$ (left
    panel) and the corresponding temperature derivatives (right panel)
    as a function of the temperature for vanishing chemical potential.
    The Polyakov-loop parameter $T_0$ is set to its pure Yang-Mills
    value $T_0=270$ MeV.}
\end{figure*}
It is left to determine the model parameters in the quark-meson
sector. As in the mean-field analysis, they are fixed to reproduce the
physical quantities in the vacuum such as the pion decay constant,
$f_{\pi} = 93$ MeV, pion and sigma meson masses, $m_{\pi} = 138$ MeV,
$m_{\sigma} = 500$ MeV, and the constituent quark mass $m_q = 298$ MeV
in the infrared. The explicit symmetry breaking term
$c= m_{\pi}^2 f_{\pi}$ is an external field and is scale independent.

\section{Phase structure}
\label{sec:phase_structure}

We first discuss the order parameters at vanishing chemical potential
and compare our findings with lattice results \cite{Bazavov:2009mi,
  Borsanyi:2010bp} as well as continuum QCD results
\cite{Braun:2009gm}. We evaluate the validity of our approach, in
particular, the need for the inclusion of the back-reaction of the
matter sector via a $N_f$ and $\mu$-dependence of the dynamical
transition parameter $T_0$ in the Polyakov-loop potential.

\Fig{fig:vev208} summarizes our findings for $\mu=0$ and the dynamical
transition parameter $T_0=208$ MeV. It shows in the left panel the
chiral condensate, normalized with the vacuum value and the
(degenerated) Polyakov-loop variable as a function of the temperature.
In the right panel, the corresponding temperature derivatives of the
order parameters are displayed. As one can see from \Fig{fig:vev208},
both widths in the temperature derivatives are comparable. For the
chiral transition a slightly larger critical temperature,
$T_c \sim 190$ MeV as for the deconfinement transition, $T_c \sim 175$
MeV, is found. This relates to the standard scenario that chiral
symmetry restoration occurs at higher temperatures than deconfinement.
We remark that the arguments for the standard scenario are strictly
valid only for phase transitions rather than for smooth crossovers as
in the present case. Moreover, the chiral and
confinement-deconfinement crossover temperatures agree within the
respective widths. The above findings compare well with a two flavor
QCD computation in the chiral limit \cite{Braun:2009gm}, and also with
recent lattice results \cite{Bazavov:2009mi, Borsanyi:2010bp}. For the
comparison with the latter one has to bear in mind that the
definitions of the chiral order parameter are ambiguous which
influences the crossover temperature.

If we switch off the back-reaction of the matter sector and stick to
the pure Yang-Mills value of the transition parameter, $T_0= 270$ MeV
the situation changes quantitatively as can be seen from
\Fig{fig:vev270}: the chiral and Polyakov derivatives peak at similar
critical temperatures about $T_c \sim 220$ MeV but the width of the
Polyakov derivative is very broad suggesting a peak substructure. This
broad structure does not compare well with the lattice predictions nor
does the high crossover temperatures. This demonstrates the importance
of the dynamics of $T_0$ already at vanishing chemical potential.

\begin{figure*}
  \subfigure{\includegraphics[width=\twofigs]{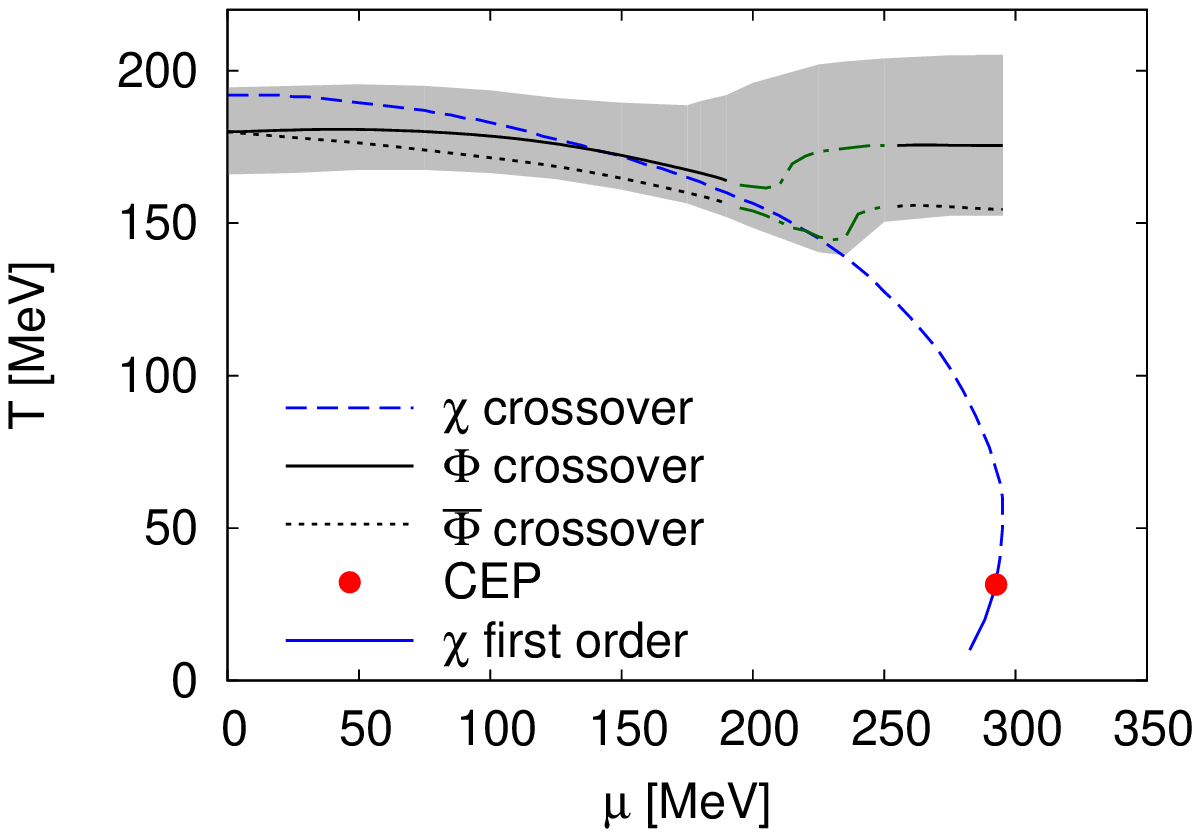}}
  \subfigure{\includegraphics[width=\twofigs]{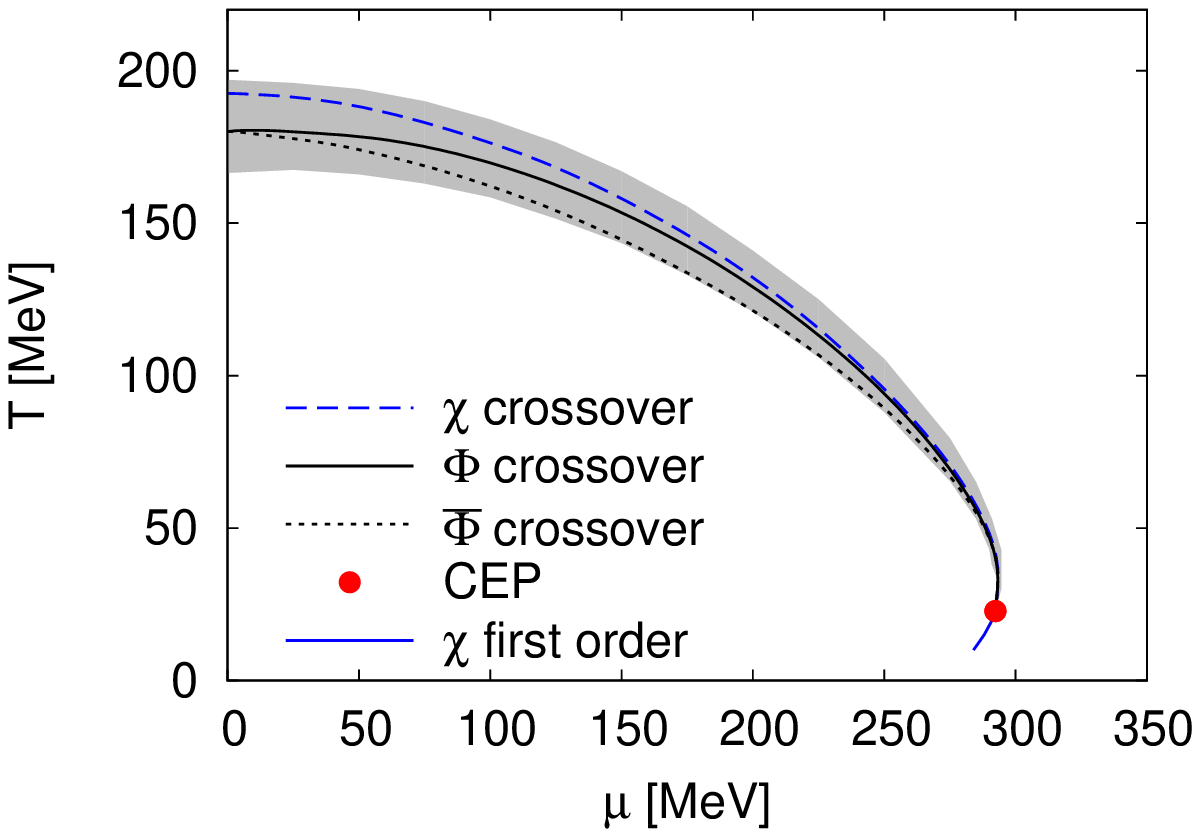}\label{fig:T0mu208g085}}
  \caption{\label{fig:phasediagram} Chiral and deconfinement phase
    diagram for a constant $\pT=208$ MeV (left panel) and for
    $T_0(\mu)$ with $\hat{\gamma}=0.85$ (right panel). The (grey) band
    corresponds to the width of $d\Phi/dT$ at $80\%$ of its peak
    height. Close to the intersection point of the chiral transition
    and the deconfinement transition at mid chemical potential a
    double peak structure in the temperature derivative of the
    Polyakov-loop variables emerges. The (green) dashed line in this
    region follows the highest peak.}
\end{figure*}

Now we extend our analysis to finite chemical potential. Our findings
are summarized in \Fig{fig:phasediagram}. The left panel shows, for
comparison, the phase diagram without back-reaction of the matter
sector to the glue sector. The right panel shows the phase diagram
with the full dynamics: the quark-meson dynamics in the presence of
the Polyakov-loop background is included with the flow \Eq{eq:flow}
whereas the back-reaction of the matter sector to the glue sector is
encoded in $T_0(N_f,\mu)$ as given in \Eq{eq:t0mu}. In the right panel
of \Fig{fig:phasediagram} we have used $\hat\gamma = 0.85$ and
$\pT(2,0) = 208$ MeV which compares well to the QCD results in
\cite{BHP} for small densities. The shaded band corresponds to the
width of $d\Phi/dT$ at $80\%$ of its peak height. We observe, that the
width of this band shrinks with increasing $\mu$. Thus, the
deconfinement transition gets sharper at higher $\mu$. Furthermore, we
can unambiguously distinguish the peak positions of the chiral and
deconfinement transition in the temperature derivatives. Even when we
vary the parameter $\hat\gamma$ almost no differences emerge. Only for
value of $\hat\gamma$ close to one an ambiguity in the peak positions
of the temperature derivatives arises, similar to the constant
$T_0$-case discussed below.

We see that the deconfinement transition line stays close to the
chiral phase boundary. It has been speculated that the quarkyonic
phase \cite{McLerran:2007qj} is signaled by a region with confinement
and chiral symmetry. Mean field computations in the PNJL and PQM model
with constant $T_0$ have supported this scenario. As has been shown in
\cite{Schaefer:2007pw}, this does not hold true if the dynamics of the
transition parameter $T_0$ is taken into account. Here we have
confirmed that in the fully dynamical PQM model the prediction in
\cite{Schaefer:2007pw} holds.

The chiral first-order transition line arising at small temperatures
and large chemical potentials terminates in a CEP which is located at
$\mu_{\text{CEP}} \sim 292$ MeV, $T_{\text{CEP}} \sim 23$ MeV. As
discussed in \cite{Schaefer:2008hk} the precise location of the CEP
depends on the chosen parameters in the vacuum, in particular on the
sigma meson mass. The back-bending of the first-order transition line
in the phase diagram towards smaller chemical potential is typical for
a FRG calculation \cite{Schaefer:2004en}. But very close to the
$\mu$-axis the slope of the first-order line tends back to infinity
similar to a mean-field treatment (not shown in the figure). This
behavior is also in agreement with the Clausius-Clapeyron relation
according to which the transition line should hit the $\mu$-axis
perpendicular.

If we switch off the back-reaction of the matter fluctuations to the
Yang-Mills sector and choose a constant $T_0=208$ MeV, the above
picture changes drastically. For finite chemical potential the
Polyakov loops $\Phi$ and $\bar\Phi$ start to deviate and the widths
of their temperature derivatives increase over the whole phase
diagram. The resulting phase diagram for a constant $T_0$ is
summarized in the left panel of \Fig{fig:phasediagram}. In the
vicinity of the intersection point of the chiral transition and the
deconfinement transition which are both smooth crossovers around
$\mu \sim 180$ MeV a double peak structure in the corresponding
temperature derivatives of the Polyakov-loop variables occur. This
hampers a unique identification of the transition point. In order to
clarify this behavior we plot the maximum of the peak location of the
$T$-derivatives in the phase diagram together with a band around
$80\%$ of the maximum value. For larger chemical potential the peak
locations deviate strongly and a coincidence of the chiral and
deconfinement transitions can be excluded. This brings back the
chirally symmetric and confined region which has been connected to the
quarkyonic phase. In summary, we have shown that this signature is
very sensitive to the correct implementation of the back-reaction of
the matter sector to the glue sector, and is most likely not present.

At high chemical potential and small temperatures, a first-order
chiral phase transition takes place which ends in a critical endpoint
of second order located at $\mu_{\text{CEP}} = 293$ MeV,
$T_{\text{CEP}} \sim 32$ MeV.
\begin{figure*}
  \subfigure{\includegraphics[width=\twofigs]{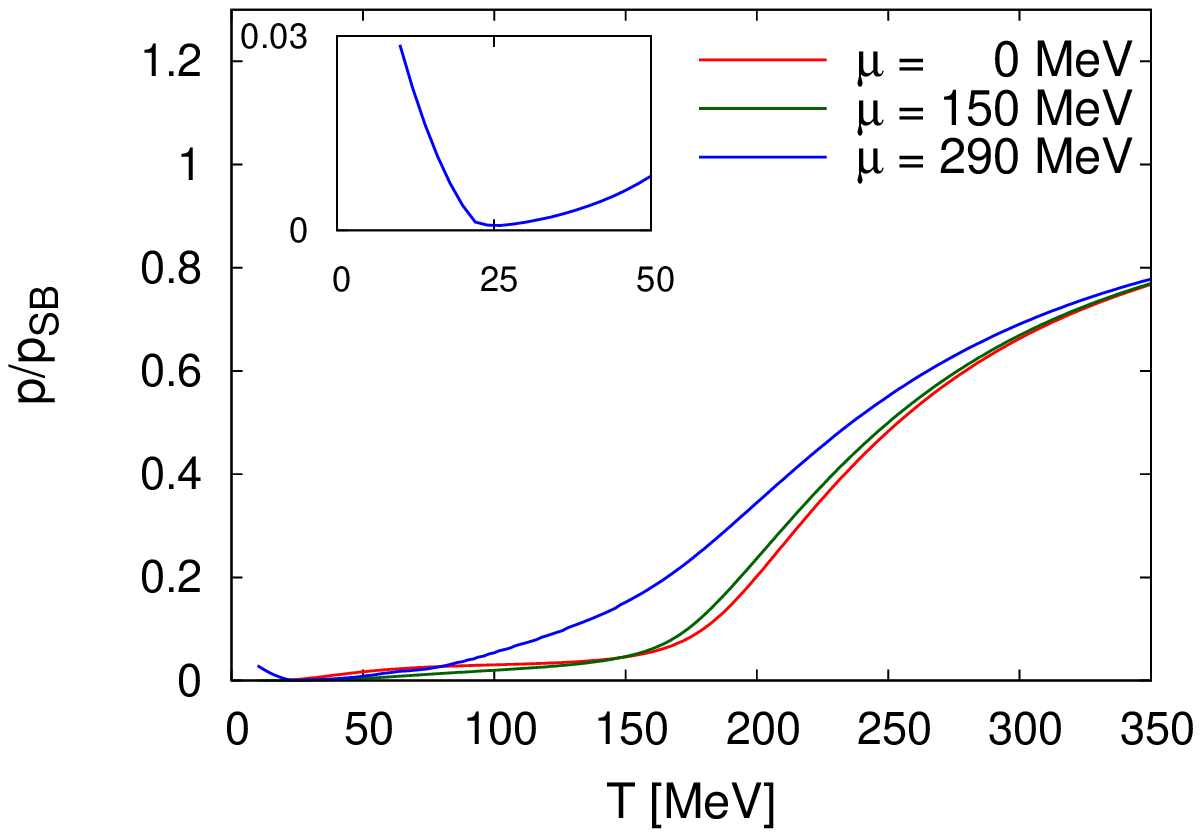}}
  \subfigure{\includegraphics[width=\twofigs]{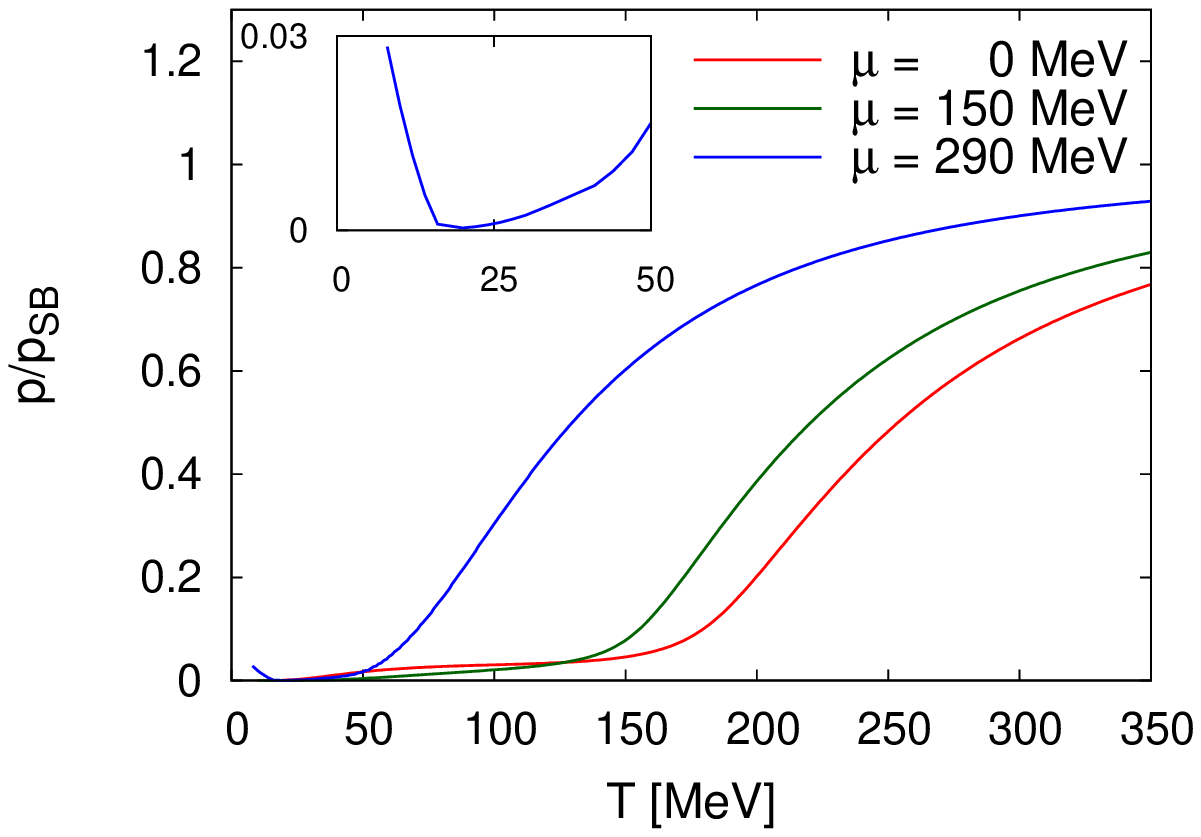}}
  \caption{\label{fig:pressure} Pressure normalized to the
    Stefan-Boltzmann pressure for a constant $\pT= 208$ MeV (left
    panel) and with $\mu$-corrections (right panel) for three
    different chemical potentials. The CEP is located approximately at
    $\mu=293$ MeV. The insets show the pressure at $\mu=290$ MeV for
    small temperatures.}
\end{figure*}

\subsection{Thermodynamics}

In general, the initial action given at the ultraviolet cutoff is
independent of the temperature and chemical potential which limits the
reliable calculation of thermodynamic quantities to a certain
temperature and density region. Due to the ultraviolet cutoff high
thermal and quantum fluctuations are suppressed. As a consequence,
cutoff-independent predictions can be obtained for temperatures
$T \lesssim \Lambda/8$ which in our case yields an upper temperature
bound of $T \sim 120$ MeV. In order to cure this cutoff remnant at
high temperature one has to combine the FRG with the expected
perturbative results. The inclusion of the missing high-momentum modes
can be achieved in an effective way by adding to the original flow,
\Eq{eq:flow}, a flow equation for an interacting Polyakov-loop quark
system for scales $k > \Lambda$. Note, that an
explicit gluon contribution to the flow equation is neglected here
because the effective Polyakov-loop potential is fitted to reproduce
the Stefan Boltzmann (SB) limit at high temperatures. In this way we
employ the equation
\begin{eqnarray}
  \label{eq:uv_omega}
  \partial_k\Omega^k_{\Lambda}(T,\mu) &=& \frac{-N_cN_f k^4}{3 \pi^2 E_q}\\
  & & \left[1-N_q(\Phi, \Phibar; T, \mu)-N_{\bar{q}}(\Phi, \Phibar; T,
    \mu)\right]\ ,\nonumber 
\end{eqnarray}
with $E_q = \sqrt{k^2 + m_q^2}$ as previously defined. This equation
is integrated from $k=\infty$ to $k=\Lambda$ and yields the
UV-contribution $\Omega_{\Lambda}^\infty(T,\mu)$, which is then added to the
grand potential resulting from the solution of the RG flow equation
(\ref{eq:flow}). The divergent zero mode contribution in
\Eq{eq:uv_omega} is neglected here. However, for vanishing quark
masses this term represents an unobservable shift in the grand
potential.

\begin{figure*}
  \subfigure{\includegraphics[width=\twofigs]{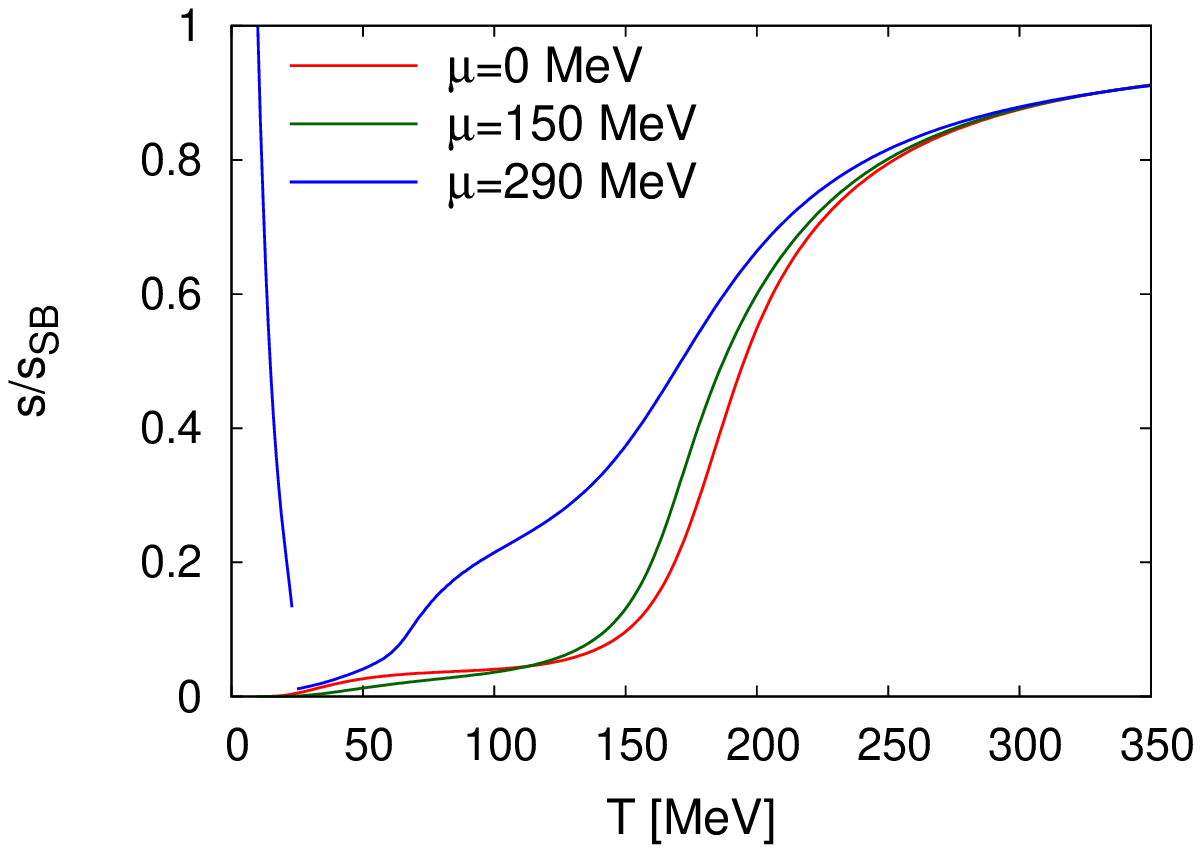}}
  \subfigure{\includegraphics[width=\twofigs]{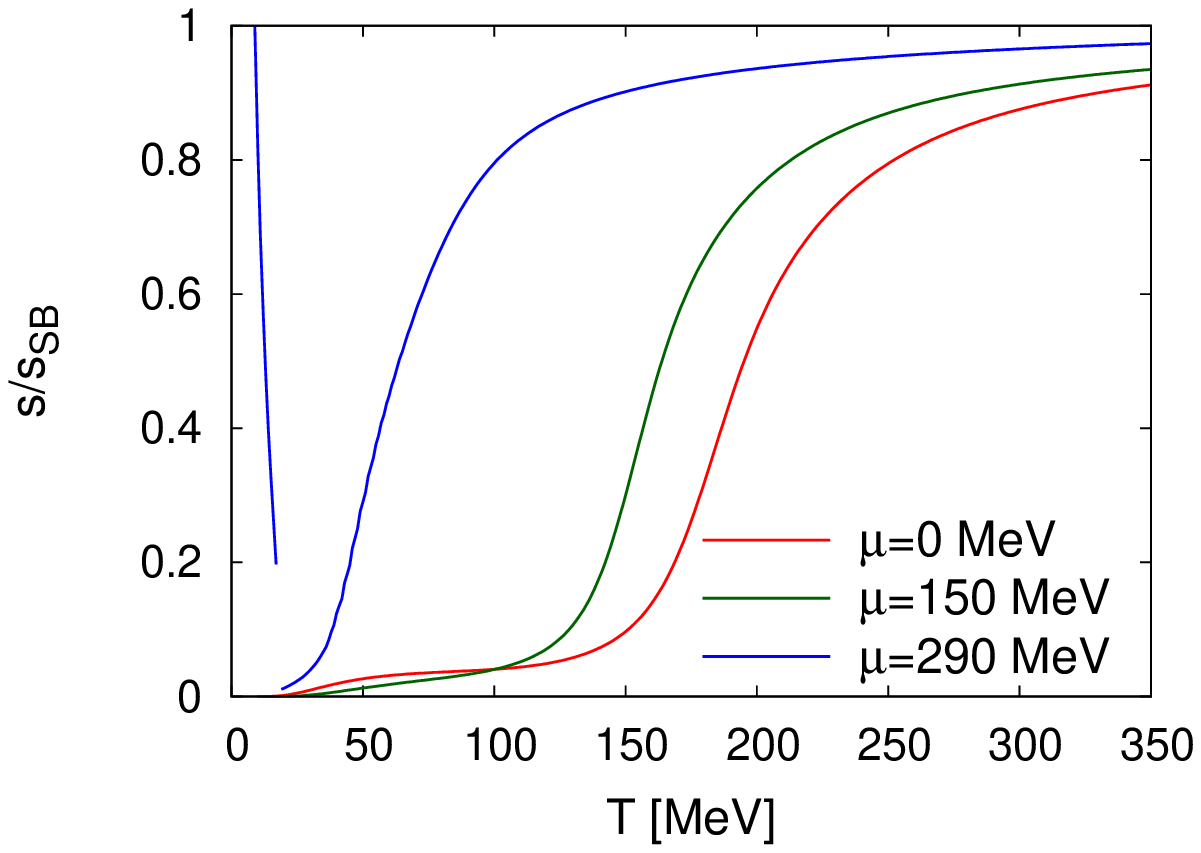}}
  \caption{\label{fig:entropy} Entropy density normalized to the
    Stefan-Boltzmann value for a constant $\pT=208$ MeV (left panel)
    and with $\mu$-corrections (right panel). At the first-order phase
    transition the entropy density jumps.}
\end{figure*}

\begin{figure*}
  \subfigure{\includegraphics[width=\twofigs]{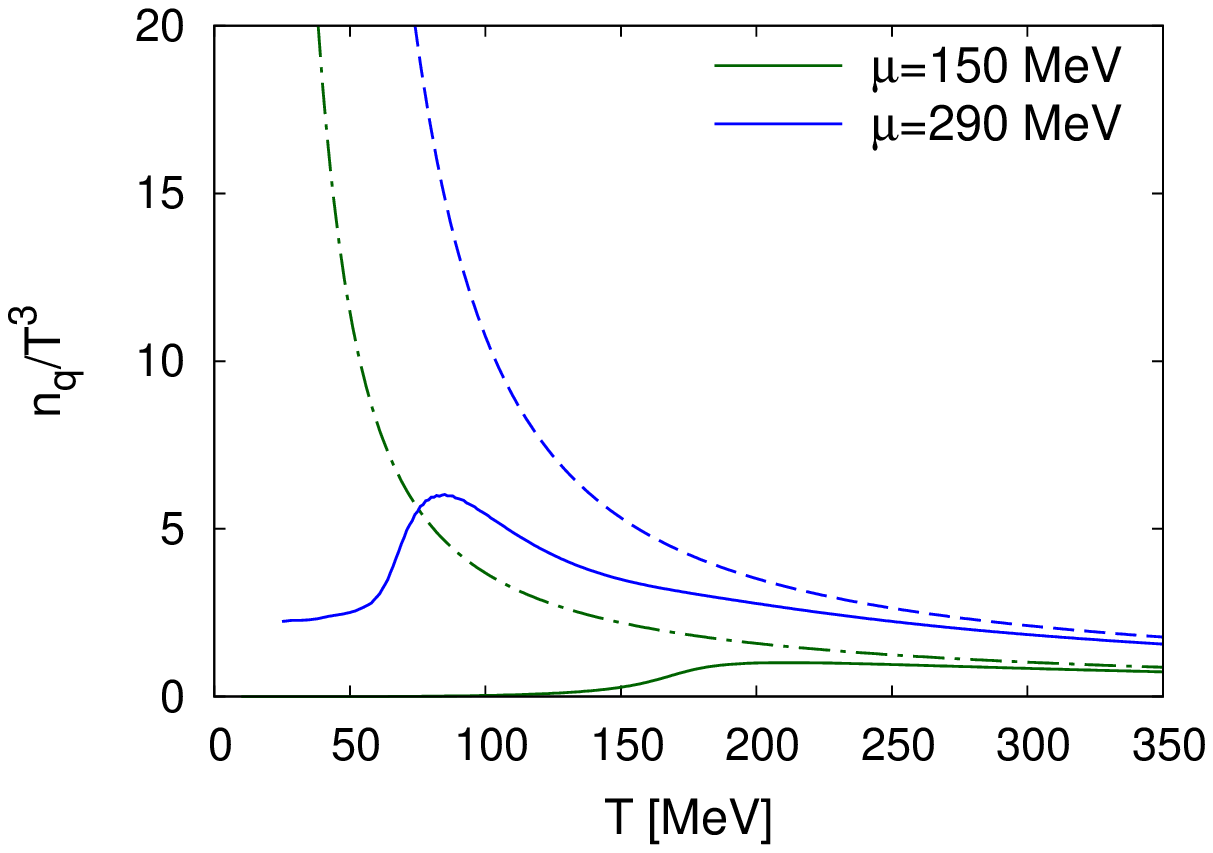}}
  \subfigure{\includegraphics[width=\twofigs]{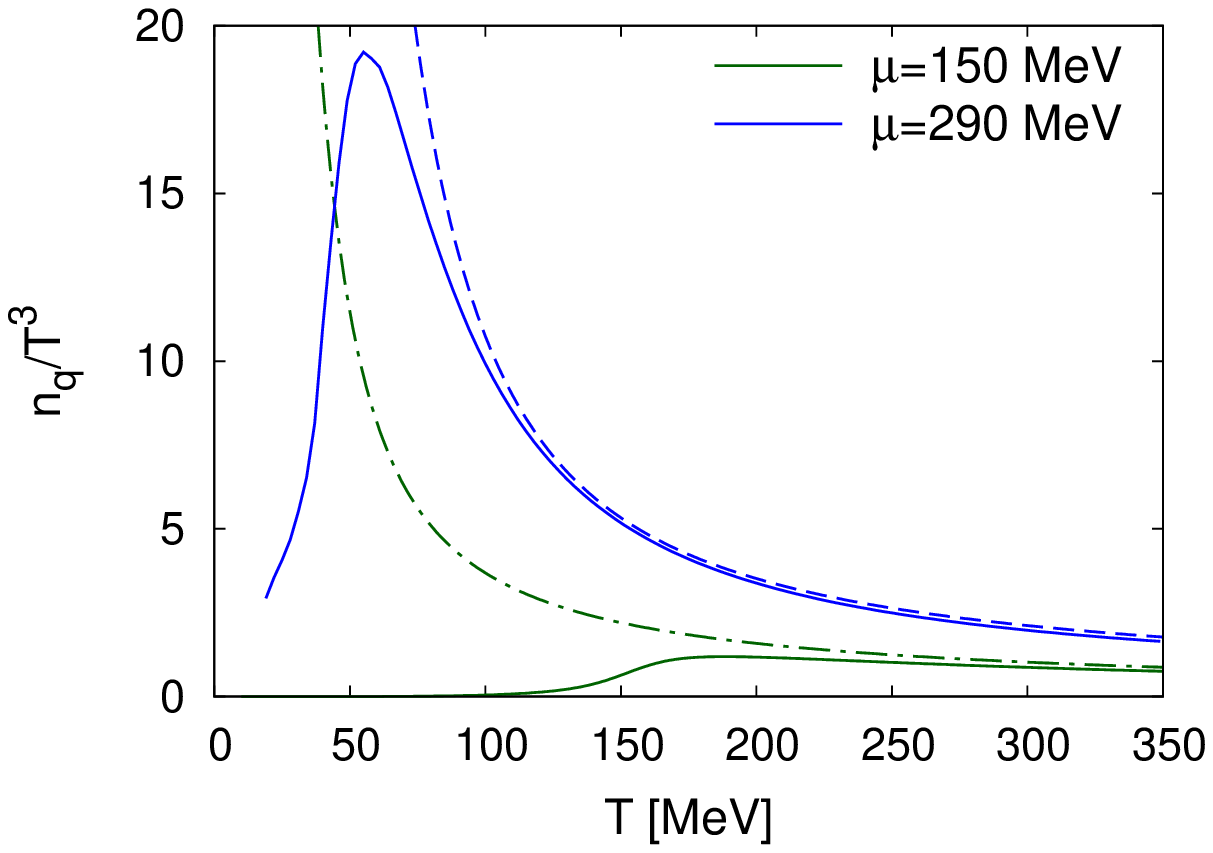}}
  \caption{\label{fig:nqdensity} Quark number density $n_q$ for two
    different chemical potentials as a function of the temperature for
    a constant $\pT=208$ MeV (left panel) and with $\mu$-corrections
    $\pT(\mu)$ (right panel). The dashed lines are the corresponding
    Stefan-Boltzmann densities. The curve for $\mu=290$ MeV starts at
    the critical temperatures of the first-order phase transition.}
\end{figure*}

In \Fig{fig:pressure} the thermodynamic pressure normalized to the
Stefan-Boltzmann pressure for three different quark chemical
potentials is shown as a function of temperature. In the left panel of
this figure, a fixed $T_0 = 208$ MeV has been used while in the right
panel the $\mu$-corrections are taken into account. The inset displays
the pressure for $\mu=290$ MeV which is close to the critical endpoint
in the phase diagram. For small temperatures the pressure decreases
and has a kink at the critical temperature due to the first-order
transition. Without the back-reaction of the matter fluctuations to
the Yang-Mills sector a similar behavior in all three curves is
observed. In the vicinity of the chiral transition the pressure
increases due to the melting of the quark masses and saturates at
about $80\%$ of the corresponding ideal gas limit which reads for
$N_f$ massless quarks and $(N_c^2 - 1)$ massless gluons
\begin{equation}
  \frac{p_\textrm{SB}}{T^4} = \frac{N_f N_c}{6} \left[\frac{7\pi^2
    }{30} + \left(\frac{\mu}{T} \right)^2 + \frac{1}{2
      \pi^2}\left(\frac{\mu}{T} \right)^4\right]+ (N_c^2 - 1)
  \frac{\pi^2 }{45} \ .  
\end{equation}
Including the back-reaction of the matter sector via the inclusion of
$T_0(\mu)$ changes the thermodynamics at larger chemical potential.
The pressure increases much faster and saturates at $\mu=290$ MeV at
$95\%$ of the Stefan-Boltzmann limit.

A similar trend is seen in the entropy, \Fig{fig:entropy}, and quark
number density, \Fig{fig:nqdensity}, if the $\mu$-corrections are
taken into account. The entropy density decreases for small
temperatures at $\mu=290$ MeV since the number of active degrees of
freedom decreases when approaching the first-order transition from
below. At the transition the entropy jumps. The bump around
$T \sim 90$ MeV (left panel) is a remnant from the smooth chiral
crossover transition. This effect is completely washed out when the
$\mu$-corrections are included (right panel). Similar to the findings
for the pressure these corrections become more significant at larger
chemical potential.
\begin{figure*}
  \includegraphics[width=\twofigs]{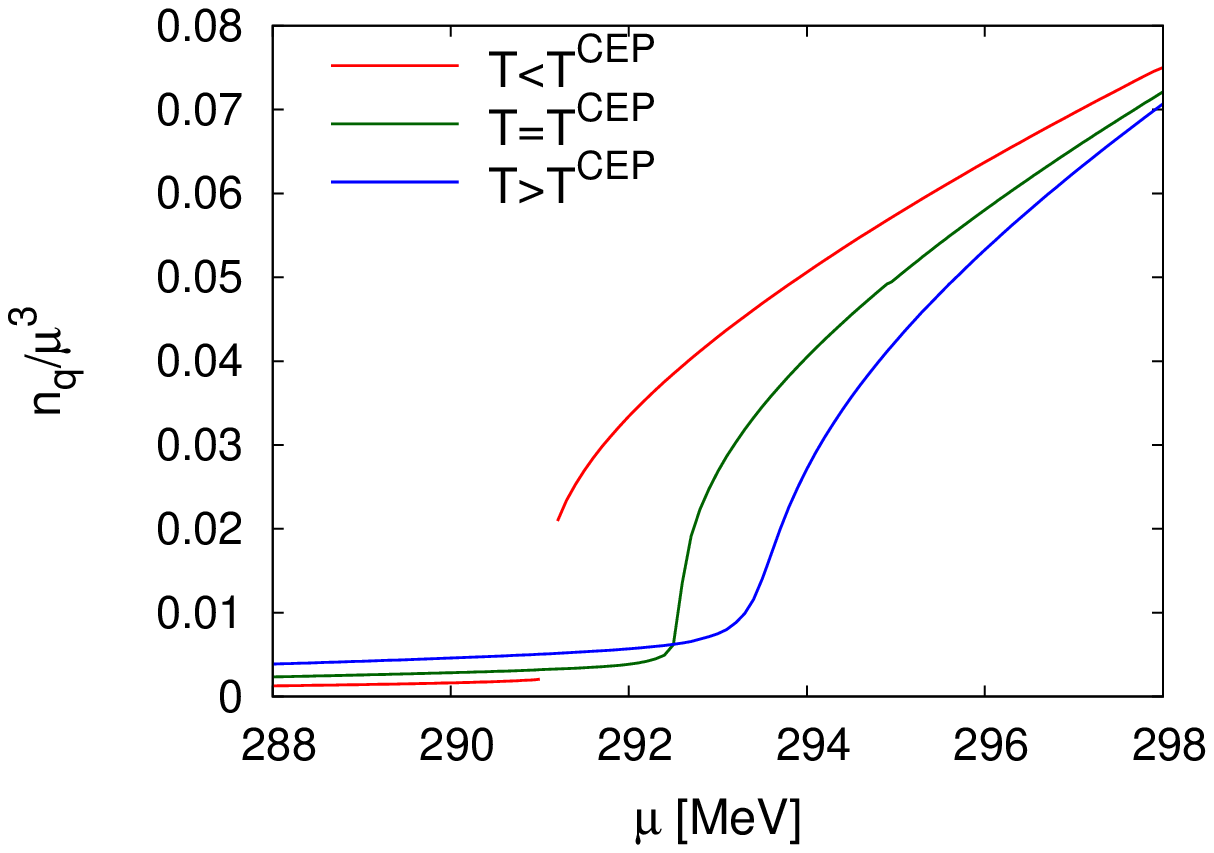}
  \includegraphics[width=\twofigs]{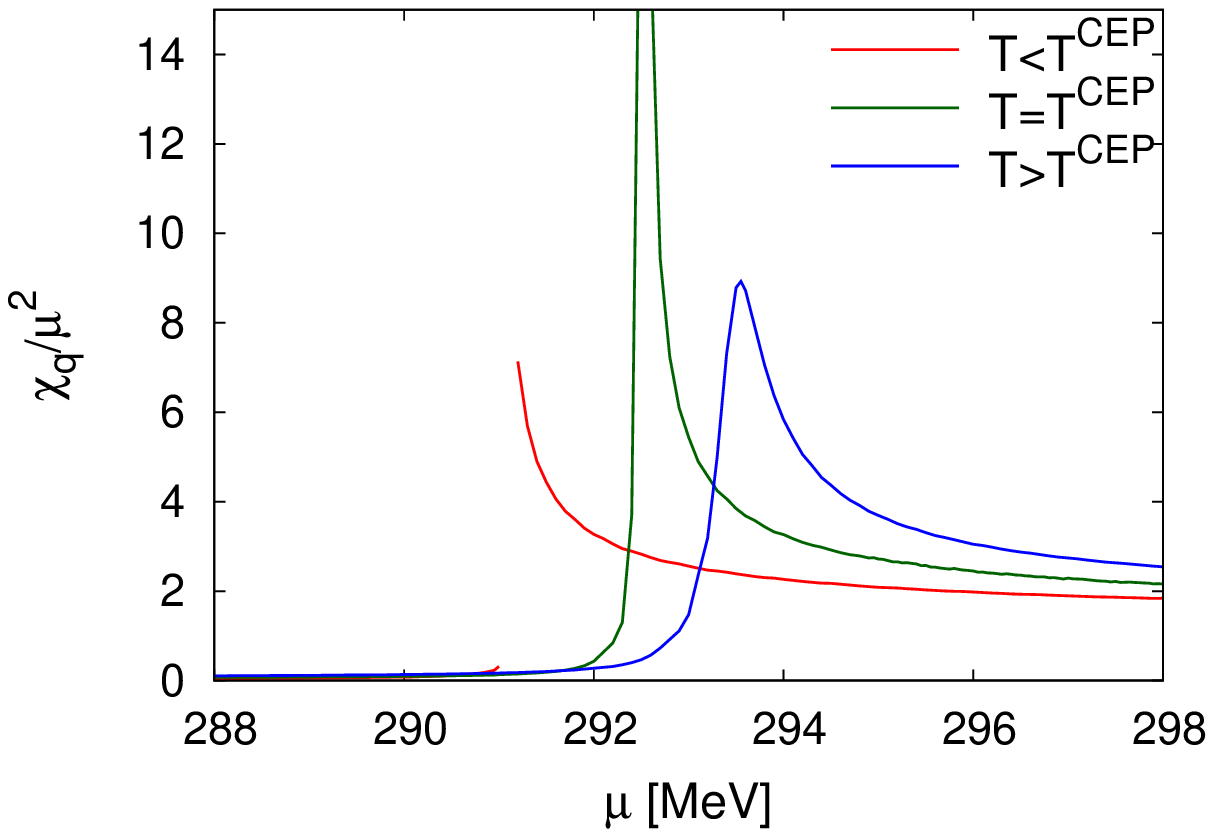}
  \caption{\label{fig:nqsus_T0const} Quark number density and the
    corresponding susceptibility in the vicinity of the CEP for three
    different temperatures. ($T=T^\textrm{CEP} \pm 5$ MeV with a constant
    $\pT=208$ MeV, $T^\textrm{CEP}\sim 32$ MeV).}
\end{figure*}
\begin{figure*}
  \includegraphics[width=\twofigs]{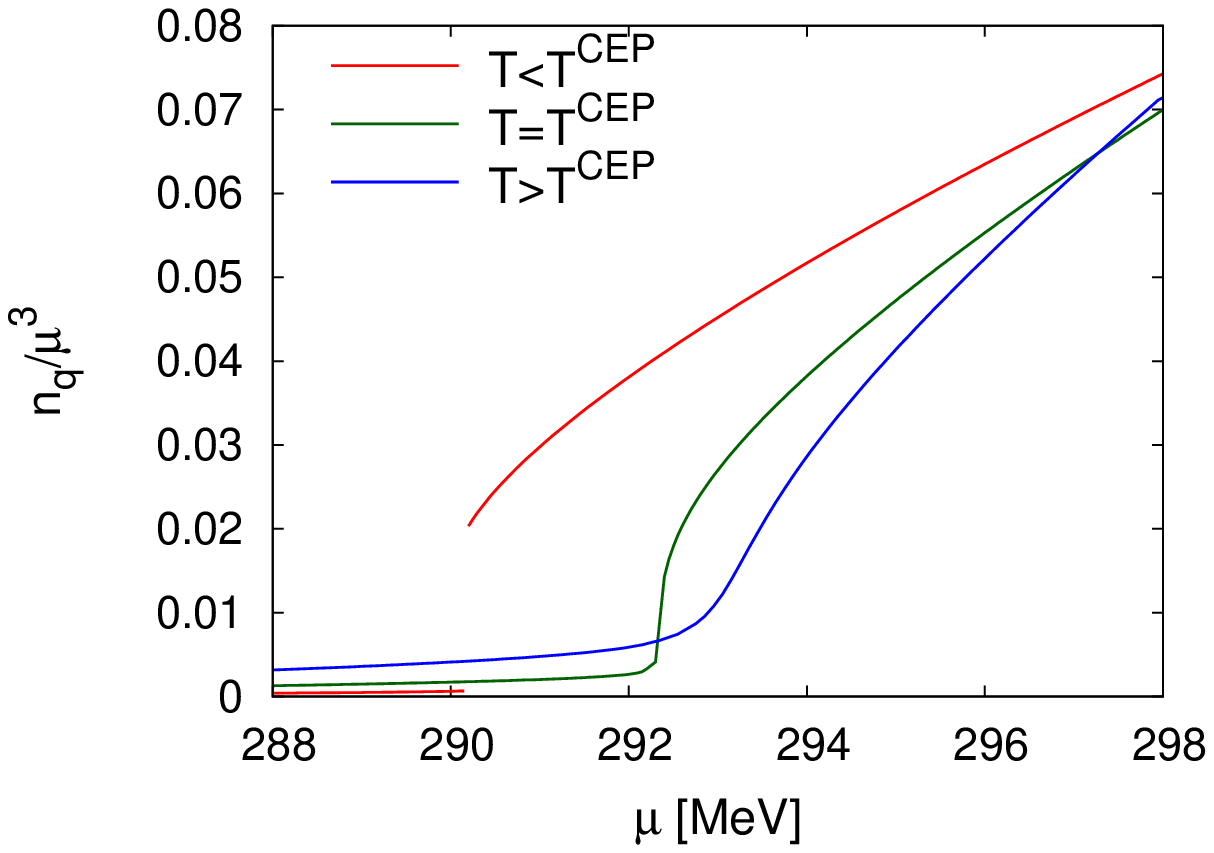}
  \includegraphics[width=\twofigs]{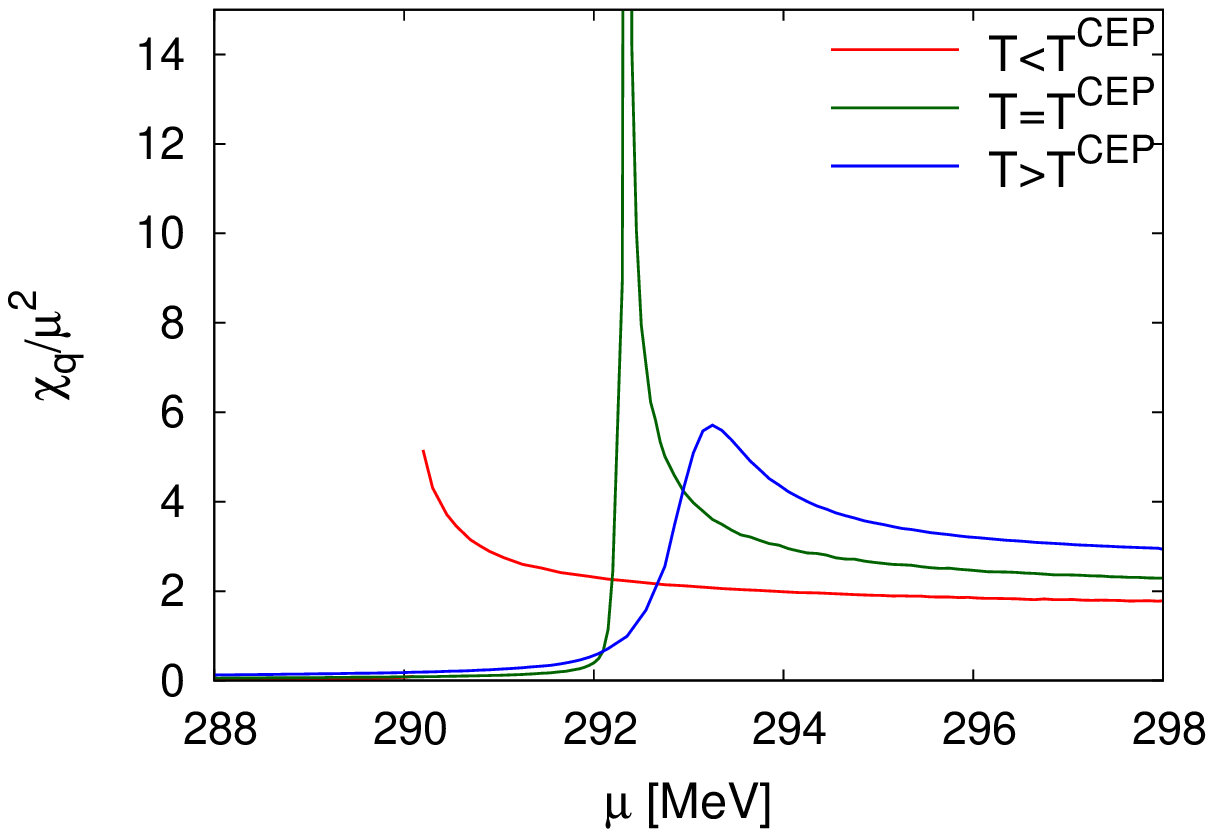}
  \caption{\label{fig:nqsus_T0} Similar to \Fig{fig:nqsus_T0const} for
    $T^\textrm{CEP}\sim 23$ MeV but with $\pT(\mu)$.}
\end{figure*}

This also appears in the quark number density $n_q = -\partial \Omega /
\partial \mu$ which is plotted in \Fig{fig:nqdensity}. For
comparison the corresponding SB-limits (dashed lines) are also shown
in this figure. The quark density approaches the SB-limit always from
below. Without the $\mu$-corrections the Polyakov loop suppresses the
quark densities for chemical potential larger than the intersection
point of the chiral and deconfinement transition in the phase diagram
\Fig{fig:phasediagram}. With the $\pT(\mu)$ corrections both
transitions coincide over the whole phase diagram and as a consequence
the quark number density approaches much faster the SB-limit (right
panel of \Fig{fig:nqdensity}).

In \Fig{fig:nqsus_T0const} the scaled quark number density (left
panel) and the corresponding scaled quark number susceptibility (right
panel) for three different temperature slices around the critical
endpoint ($T^\textrm{CEP}, T^\textrm{CEP} \pm 5$ MeV) as a function of
the quark chemical potential are collected. In this figure the
$\mu$-corrections in $\pT$ are omitted while in \Fig{fig:nqsus_T0}
they are taken into account. Due to the chiral critical endpoint which
is a second-order transition the susceptibility diverges with a
certain power law \cite{Stephanov:2007fk}. There are no strong
modifications in the structure of the susceptibility divergence if the
back-reaction of the matter sector is taken into account or not. As a
consequence it seems that the size of the critical region around the
CEP is not strongly modified by these fluctuations. The only
difference is that including the $\mu$-corrections the peak height of
the susceptibility is more pronounced towards the CEP.

\section{Summary and conclusions}
\label{sec:conclusion}

In the present work we have studied the Polyakov-extended quark-meson
model beyond mean-field approximation. The quark-meson fluctuations to
the matter sector are included within a functional renormalization
group approach. In turn, the quark-meson fluctuations to the gluonic
sector are estimated by a perturbative computation as suggested in
\cite{Schaefer:2007pw}. Interestingly, the latter procedure has been
recently confirmed by a full dynamical QCD computation
\cite{Braun:2009gm, BHP}.

The validity range of the present model includes the temperature
regime about the crossover temperature $T_c\approx 200$ MeV up to
medium quark chemical potential of $\mu_{q} \approx 100-200$ MeV. At
low temperature and large chemical potential the model suffers from
missing inclusion of baryonic degrees of freedom and further
resonances.

However, the quantum fluctuations lead to strong modifications of the
phase structure. Such modifications have already been observed in FRG
studies of the quark-meson model. One of the prominent effects is a
shrinking of the size of the critical region around the critical
endpoint in the phase diagram~\cite{Schaefer2006a}. Furthermore,
quantum fluctuations push the potential critical endpoint towards
lower temperatures and larger chemical potential, see,
e.g.,~\cite{Nakano:2009ps}.

Our computation entails a small likelihood for a critical point
with $\mu_{\rm B}/T\approx 1-2$ as predicted by some recent lattice
studies \cite{Gupta:2009mu}. Indeed, this is in accordance with the
arguments of \cite{deForcrand:2008zi} as well as with recent
developments concerning the convergence of the Taylor expansion about
vanishing chemical potential \cite{karsch2010}. The latter also
complicates the extraction of the location of the critical point from
the analysis of higher moments such as investigated in
\cite{Athanasiou:2010kw}. Thus, an extension of the present
computations beyond mean field towards the low temperature high
density regime will provide valuable information. 

Further results concern thermodynamical quantities such as the
pressure and the density. At vanishing density these quantities agree
well with related lattice predictions and, in particular, show the
correct behavior in the transition regime from mesonic degrees of
freedom to the quark-gluon plasma regime.

In a forthcoming publication \cite{HPSW} we will present a full
solution including novel algorithmic differentiation techniques
\cite{Wagner:2009pm} for the flow \Eq{eq:flow} which also allows us to
discuss the systematics of the present approach towards QCD, as well
as studying further thermodynamic quantities. We also plan to extend
the present study to the $2+1$ flavor case, as well as tightening the
relation to full dynamical QCD.

\subsection*{Acknowledgments}
TKH is recipient of a DOC-fFORTE-fellowship of the Austrian Academy of
Sciences and supported by the FWF doctoral program DK-W1203-N08. JMP
acknowledges support by Helmholtz Alliance HA216/EMMI. We thank Jens
Braun, Lisa M.~Haas and Mario Mitter for discussions. We also thank
Mathias Wagner for discussions and work on related subjects.

\end{document}